\documentclass[letter,11pt]{article}
\pdfoutput=1
\usepackage{jheppub}
\usepackage[dvipsnames,table,xcdraw]{xcolor}
\usepackage[T1]{fontenc}
\usepackage[colorlinks=true, linkcolor=blue, urlcolor=blue, citecolor=blue, anchorcolor=blue]{hyperref}
\usepackage{subcaption}
\usepackage{multirow,makecell}
\usepackage{comment}

\usepackage{orcidlink}

\newcommand{\bec}{\begin{center}}
\newcommand{\eec}{\end{center}}
\newcommand{\beq}{\begin{equation}}
\newcommand{\eeq}{\end{equation}}
\newcommand{\bea}{\begin{eqnarray}}
\newcommand{\eea}{\end{eqnarray}}

\title{Thermodynamic Diagnostics for Complex Langevin Simulations: The Role of Configurational Temperature}

\author[a]{Anosh Joseph~\orcidlink{0000-0003-4288-8207},}
\author[b]{Arpith Kumar~\orcidlink{0000-0002-5887-3803}}

\affiliation[a]{National Institute for Theoretical and Computational Sciences, \\ School of Physics, and Mandelstam Institute for Theoretical Physics,\\ University of the Witwatersrand, Johannesburg, Wits 2050, South Africa}
\affiliation[b]{Key Laboratory of Quark and Lepton Physics (MOE) and Institute of Particle Physics, \\
Central China Normal University, Wuhan 430079, China}

\emailAdd{anosh.joseph@wits.ac.za}
\emailAdd{arpithk@ccnu.edu.cn}

\abstract{
The complex Langevin method (CLM) offers a potential solution to the sign problem in quantum field theories with complex actions, but can converge to incorrect results even when simulations appear stable. Existing diagnostics monitor drift distributions or Langevin-time operators but do not explicitly test whether configurations are sampled with the correct Boltzmann statistical weight. We propose a complementary diagnostic based on configurational temperature, constructed from gradients and Hessians of the action. Testing in one-dimensional PT-symmetric models demonstrates 0.2-3\% accuracy in reproducing the expected value for the configurational temperature. Crucially, configurational temperature detects algorithmic errors -- including noise mis-scaling, step-size artifacts, and incomplete thermalization -- significantly more sensitively than existing drift-based or operator-based criteria. The method relies on the derivatives of the local action, making it applicable to general lattice theories, though computational cost requires consideration in higher dimensions. Our results suggest configurational temperature as a valuable addition to CLM diagnostics,  complementing existing tools with potential applications from supersymmetric matrix models to lattice QCD at finite density.
}

\begin{document}
\maketitle
\flushbottom

\section{Introduction} 
\label{sec:intro}

We can systematically study the nonperturbative aspects of quantum field theories (QFTs) through lattice regularizations of the path integral. 
Monte Carlo methods provide a robust framework for extracting physical observables in such systems, where the central idea is to generate field configurations with probabilities proportional to the Euclidean weight, $e^{-{\cal S}}$, and to evaluate observables by statistical averaging over this importance-sampled ensemble.

Difficulties arise, however, when the action becomes complex. 
In such cases, the weight factor can no longer be interpreted as a probability distribution, giving rise to the so-called {\it sign problem}. 
This issue appears in many theories of broad physical interest: QCD at finite baryon density, where the fermion determinant acquires a complex phase; QCD with a topological $\theta$-term; Chern-Simons gauge theories with complex couplings; and certain classes of chiral gauge theories. 
The inability to directly apply standard Monte Carlo importance sampling in these contexts hampers our ability to probe their nonperturbative dynamics.

The complex Langevin method (CLM)~\cite{Klauder:1983nn, Klauder:1983zm, Klauder:1983sp, Parisi:1984cs, Damgaard:1987rr} was proposed as a strategy to circumvent the sign problem. 
The method extends the framework of stochastic quantization, formulated initially for real actions, to systems with complex actions. 
In practice, this involves a complexification of the original dynamical variables of the path integral and the introduction of a stochastic process governed by a Langevin equation with complex drift. 
Expectation values of observables in the original path integral are then recovered as equilibrium averages over this stochastic process.
For recent reviews of the method, see~\cite{Berger:2019odf, Joseph:2025tfw}.

CLM has been applied successfully to a variety of models~\cite{Berges:2005yt, Berges:2006xc, Berges:2007nr, Bloch:2017sex, Aarts:2008rr, Pehlevan:2007eq, Aarts:2008wh, Aarts:2009hn, Aarts:2010gr, Aarts:2011zn}, including relativistic Bose gases at finite chemical potential, low-dimensional QCD-like systems, and spin models at nonzero density. 
It has also been employed in the study of supersymmetric matrix models~\cite{Ito:2016efb, Ito:2016hlj, Anagnostopoulos:2017gos, Joseph:2019sof, Joseph:2020gdh, Kumar:2022fas, Kumar:2022giw, Kumar:2023nya}, and in large-$N$ unitary matrix models where it has been used to explore Gross--Witten--Wadia transitions~\cite{Basu:2018dtm, Gross:1980he, Wadia:2012fr, Wadia:1980cp}. 
Complex Langevin simulations have already been applied to QCD at finite density in several works, beginning with the pioneering simulations of Sexty \cite{Sexty:2013ica}, followed by investigations by multiple groups studying thermodynamic observables and the equation of state \cite{Aarts:2014bwa, Fodor:2015doa, Nagata:2018mkb, Sexty:2019vqx, Kogut:2019qmi, Ito:2020mys}.

A recent analysis of the necessary and sufficient conditions for the correctness of complex Langevin simulations was presented in Ref.~\cite{Mandl:2025mav}. It was shown that, if in a given theory the expectation values of all observables within a suitable functional space satisfy the corresponding Schwinger--Dyson equations and certain boundedness criteria, these expectation values are guaranteed to be correct. Less rigorous, yet foundational, conditions were previously discussed in Ref.~\cite{Seiler:2023kes}, which served as the basis for Ref.~\cite{Mandl:2025mav}. Related investigations can be found in Refs.~\cite{Scherzer:2018hid,Scherzer:2019lrh}.

Despite these advances, it is well known that CLM does not always converge to the correct result: simulations can appear stable while still producing systematically wrong answers. 
This has motivated the development of practical reliability and diagnostics criteria, such as monitoring the distribution of the drift term or checking the action of the Langevin-time evolution operator on observables. 
While useful, these diagnostics are insufficient, particularly in models with many degrees of freedom or subtle algorithmic instabilities. 
In this work, we propose an alternative and complementary diagnostic criterion based on the idea of configurational temperature, derived from statistical mechanics.
This criterion directly tests the sampling consistency in reaching the target equilibrium distribution.

This paper is organized as follows. 
In Sec.~\ref{sec:Complex_Langevin_Method_and_Its_Limitations} we briefly review the complex Langevin method and summarize its main limitations. 
In Sec.~\ref{sec:Configuration-Based_Thermometer} we introduce the concept of a configuration based thermometer, while Sec.~\ref{sec:A_brief_derivation_of_the_configuration-based_thermometer} provides a brief derivation of the explicit form of the configurational temperature estimator and Sec.~\ref{sec:Application_to_Euclidean_lattice_field_theory} points out the connection to Euclidean lattice field theory. 
We demonstrate the effectiveness of this estimator in one-dimensional PT-symmetric models in Sec.~\ref{sec:One-dimensional_PT-symmetric_theories}, where we compare it against existing correctness criteria. 
Sec.~\ref{sec:Numerical_Tests_of_the_Estimator} discusses the numerical tests of the estimator such as detecting algorithmic errors, dependence of Langevin step size, and monitoring thermalization.
In Sec.~\ref{sec:Comparison_with_Existing_Diagnostics} we discuss the comparison with existing diagnostics and computational cost of the estimator and scalability.
Finally, in Sec.~\ref{sec:Discussion_and_Outlook}, we summarize our findings and outline potential applications to higher-dimensional scalar and gauge theories, with an eye toward future applications in lattice QCD at finite density.

\section{Complex Langevin Method and Its Limitations} 
\label{sec:Complex_Langevin_Method_and_Its_Limitations}

The central idea of stochastic quantization is that expectation values of observables can be obtained as equilibrium values of a stochastic process. 
In Langevin dynamics, this is realized by evolving the system in a fictitious time direction, $\theta$, under the influence of stochastic noise. 
When the action is complex, one can attempt to extend this framework, although in such cases the field variables necessarily become complexified during the Langevin evolution, since the gradient of the action, the {\it drift term}, is itself complex.

For a field theory with action ${\cal S}[\phi]$, the complex Langevin equation in Euler discretized form reads
\beq
\phi (\theta + \Delta \theta) 
= \phi (\theta) - \Delta \theta \, \frac{\delta {\cal S}[\phi]}{\delta \phi (\theta)} + \sqrt{\Delta \theta} \, \eta (\theta),
\eeq
where $\Delta \theta$ is the Langevin time step, and $\eta(\theta)$ is a Gaussian noise satisfying
\beq
\label{eq:Gaussian_noise}
\langle \eta (\theta) \rangle = 0, 
\qquad 
\langle \eta (\theta) \eta (\theta') \rangle = 2 \delta_{\theta,\theta'}.
\eeq
In practice, real Gaussian noise is often employed to suppress large excursions in the imaginary directions of the fields~\cite{Aarts:2009uq, Aarts:2011ax, Nagata:2015uga}.

For a general operator $\mathcal{O}$, the noise-averaged expectation value is
\beq
\big\langle \mathcal{O} [\phi(\theta)] \big\rangle_\eta = \int d\phi \, P[\phi(\theta)] \, \mathcal{O}[\phi],
\eeq
where the probability distribution $P[\phi(\theta)]$ evolves according to the Fokker--Planck equation
\beq
\frac{\partial P[\phi(\theta)]}{\partial \theta} = \frac{\delta}{\delta \phi(\theta)} 
\left( \frac{\delta}{\delta \phi(\theta)} + \frac{\delta {\cal S}[\phi]}{\delta \phi(\theta)} \right) 
P[\phi(\theta)].
\eeq
For real actions, the stationary solution is 
\beq
P[\phi] \propto e^{-{\cal S}[\phi]}, \nonumber
\eeq
ensuring convergence to the correct equilibrium distribution as $\theta \to \infty$.  
For complex actions, however, the situation is far more subtle: the drift is complex and the fields evolve into a complexified manifold $\phi = \mathrm{Re}\,\phi + i \,\mathrm{Im}\,\phi$. Rigorous proofs of convergence to the desired complex measure remain elusive~\cite{Parisi:1984cs, Klauder:1985kq, Klauder:1985ks, Gausterer:1986gk}.

It has long been known that while Langevin dynamics with real actions generally yield correct results, the complex Langevin method can converge to incorrect limits. 
Two main obstructions have been identified.  
First, the {\it excursion problem}: complexified variables can drift far into the imaginary directions, invalidating the integration by parts argument underlying the method~\cite{Aarts:2009uq, Aarts:2011ax}.  
Second, the {\it singular drift problem}: if the drift term has poles, the dynamics may approach them frequently, spoiling the proof of correctness~\cite{Nishimura:2015pba}.  
These insights clarified why the complex Langevin method may appear stable yet yield wrong answers.

As a result, much effort has gone into formulating practical correctness criteria.  
One proposal~\cite{Aarts:2009uq} introduces the Langevin-time evolution operator $L$ acting on observables, and demands $\langle L \mathcal{O} \rangle = 0$ in the long-time limit.  
While useful in simple models, this suffers from large statistical fluctuations in systems with many degrees of freedom and cannot guarantee correctness.  
A sharper criterion was later proposed by Nagata {\it et al.}~\cite{Nagata:2016vkn}, who showed that correct convergence requires the probability distribution of the drift term to decay exponentially (or faster) at large magnitude.  
This condition is necessary and sufficient, provided additional assumptions such as ergodicity are met~\cite{Aarts:2017vrv, Seiler:2017wvd}.  
Since the drift is computed at every Langevin step, this diagnostic is inexpensive to monitor.

Current reliability criteria for CLM may be challenging to apply in large or complicated theories.  
This motivates the search for complementary diagnostics more closely tied to physical observables.  
In this work, we propose an alternative: a configurational temperature estimator constructed from the gradient and Hessian of the complex action.  
As we will demonstrate, this criterion directly probes sampling consistency and provides a sensitive and independent cross-check of CLM simulations.

\section{Configuration-Based Thermometer}
\label{sec:Configuration-Based_Thermometer}

In molecular dynamics simulations, conserved quantities such as energy or momentum often serve as valuable indicators of algorithmic correctness. 
While conservation does not by itself guarantee validity, violations typically signal programming or numerical errors that are straightforward to identify.

In canonical Monte Carlo simulations, such exact conservation laws do not exist. This makes the validation more challenging. 
Traditionally, one can assess the accuracy by comparing measured observables with known thermodynamic benchmarks. 
However, this approach becomes impractical when exploring new theories or parameter regimes where no reference data exist.

Rugh provided a breakthrough by deriving a geometric expression for the temperature in the microcanonical ensemble~\cite{PhysRevLett.78.772}. 
His formulation relates temperature to the curvature of constant-energy hypersurfaces in phase space, thereby offering a purely dynamical definition. 
Building on this idea, Butler {\it et al.} introduced a \emph{configurational temperature} suitable for canonical ensembles~\cite{10.1063/1.477301}. 
This estimator depends only on gradients and curvatures of the potential (or action), making it especially attractive for Monte Carlo simulations that do not sample momenta.
The conceptual foundations can be traced back to Landau and Lifshitz's {\it Statistical Physics}~\cite{Landau1952}, Tolman's {\it Principles of Statistical Mechanics}~\cite{Tolman1979}, and later clarified in historical accounts such as~\cite{Hoover2007}.

Subsequent studies established the generality and robustness of the approach. 
Jepps {\it et al.} showed that arbitrary phase-space vector fields can yield valid temperature estimators under suitable conditions, with supporting molecular dynamics simulations~\cite{PhysRevE.62.4757}. 
Numerical tests confirmed its accuracy in diverse settings, including Lennard--Jones fluids~\cite{10.1063/1.477301} and the two-dimensional XY model using cluster algorithms~\cite{PhysRevE.94.062113}. 
One can find broader theoretical extensions and further molecular dynamics applications in~\cite{1998JPhA...31.7761R, PhysRevE.62.4757, 10.1063/1.1348024, 10.1063/1.480995}.  

Although initially proposed for thermostat design and system control, the configurational temperature was quickly recognized as a powerful diagnostic tool. 
Because it probes thermodynamic consistency using only configurational information, it can reveal sampling inconsistencies or numerical instabilities independently of momentum-based definitions.

\subsection{A brief derivation of the configuration-based thermometer}
\label{sec:A_brief_derivation_of_the_configuration-based_thermometer}

The starting point is the thermodynamic relation
\begin{equation}
\frac{1}{T} = \left( \frac{\partial S}{\partial E} \right)_{V},
\label{eq:dE}
\end{equation}
which identifies inverse temperature with the change of entropy under an infinitesimal energy shift at fixed volume.  
In the microcanonical ensemble, the entropy is proportional to the phase-space volume enclosed by a constant-energy shell,
\begin{equation}
S(E) = k_B \ln \Omega_\Gamma(E) , \qquad 
\Omega_\Gamma(E) = \int_{\mu C(E)} d\vec{\Gamma},
\end{equation}
where $\vec{\Gamma} = (q_1,\dots,q_N; p_1,\dots,p_N)$ denotes positions and momenta, such that, $\mu C(E)\equiv \{\vec{\Gamma} \mid H(\vec{\Gamma} )\leq E\}$ for Hamiltonian $H(\vec{\Gamma} )$.

Following Rugh's geometric construction~\cite{PhysRevLett.78.772}, one introduces a displacement of each configuration along a phase-space vector field $\vec{n}(\vec{\Gamma})$ that increases the energy by $\Delta E$.  
Choosing
\begin{equation}
\vec{n}(\vec{\Gamma}) = 
\frac{\vec{\nabla}_{\vec{q}} H}{\vec{\nabla}_{\vec{q}} H \cdot \vec{\nabla}_{\vec{q}} H},
\label{eq:nvector_Gamma}
\end{equation}
ensures, to leading order, that $H(\vec{q} + \Delta E \, \vec{n}) = H(\vec{q}) + \Delta E$.  
The associated Jacobian yields the entropy shift
\begin{eqnarray}
\Delta S &=& S(E + \Delta E) - S(E) \nonumber \\
&=& k_B \ln\left( 1 + \Delta E \left\langle \vec{\nabla}_{\vec{q}} \cdot \vec{n}(\vec{q}) \right \rangle \right) \nonumber \\
&\approx& k_B \, \Delta E \, \left \langle \vec{\nabla}_{\vec{q}} \cdot \vec{n}(\vec{q}) \right \rangle,
\end{eqnarray}
where $\langle \cdots \rangle$ denotes the microcanonical average and we have used $\ln(1 + x) \simeq x$ for small $x$. 
In the limit $\Delta E \to 0$, we obtain
\begin{equation}
\frac{1}{k_B T} = \frac{\partial S}{\partial E} = \big\langle \vec{\nabla}_{\vec{q}} \cdot \vec{n}(\vec{q}) \big\rangle.
\end{equation}

For simulations without explicit momenta (as in canonical Monte Carlo or Langevin), one can set $H(\vec{q}) = \Phi(\vec{q})$, the potential energy or action.  
This leads to the configurational temperature estimator (setting $k_B = 1$)
\begin{equation}
\frac{1}{T} = \left\langle \vec{\nabla}_{\vec{q}} \cdot \frac{\vec{\nabla}_{\vec{q}} \Phi}{|\vec{\nabla}_{\vec{q}} \Phi|^2} \right\rangle + \mathcal{O}(1/N),
\label{eq:temp_eq}
\end{equation}
which is valid in the canonical ensemble by ensemble equivalence in the thermodynamic limit $N\to \infty$.

It is convenient to rewrite this in terms of the gradient and Hessian of $\Phi$.  
Let us define
\begin{equation}
\vec{g} \equiv \vec{\nabla}_{\vec{q}} \Phi, \qquad
\mathbb{H} \equiv \vec{\nabla}_{\vec{q}} \vec{\nabla}_{\vec{q}}^T \Phi.
\end{equation}
Then
\begin{equation}
\frac{1}{T} = \vec{\nabla}_{\vec{q}} \cdot 
\left( \frac{\vec{g}}{|\vec{g}|^2} \right) = \frac{\mathrm{Tr}(\mathbb{H})}{|\vec{g}|^2} 
- 2 \frac{\vec{g}^T \mathbb{H} \vec{g}}{|\vec{g}|^4}.
\label{eq:hessian-form}
\end{equation}

The physical interpretation of the two terms is instructive: the trace term measures isotropic curvature, weighted by gradient magnitude, and it dominates in high dimensions.
The directional term measures curvature along the gradient direction; it captures anisotropic effects in low dimensions.
This decomposition explains why the estimator is more sensitive in low dimensions, where both terms contribute comparably.
Equation~\eqref{eq:hessian-form} is the compact and numerically practical form of the configurational temperature estimator: it depends only on local derivatives of the action.

\subsection{Application to Euclidean lattice field theory}
\label{sec:Application_to_Euclidean_lattice_field_theory}

While the configurational temperature was originally derived for classical statistical mechanics, the Euclidean path integral formalism provides a formal bridge to quantum field theory through a mathematical analogy. 
In the path integral quantization, expectation values are computed as
\begin{equation}
\langle O \rangle = \frac{\int \mathcal{D} \phi \, O[\phi] \, e^{-{\cal S}[\phi]}}{\int \mathcal{D} \phi \, e^{-{\cal S}[\phi]}},
\label{eq:path_integral}
\end{equation}
which has the identical mathematical structure to a canonical ensemble with ``potential'' ${\cal S}[\phi]$ and ``inverse temperature'' unity:
\begin{equation}
\text{Classical:} \quad P(\vec{q}) \propto e^{-\Phi(\vec{q})/T}, \qquad
\text{Quantum:} \quad P[\phi] \propto e^{-{\cal S}[\phi]}.
\label{eq:analogy}
\end{equation}
This structural equivalence justifies the application of the configurational temperature estimator to lattice field theory as a diagnostic tool, even though the physical interpretation differs between the two contexts.

\subsubsection{Sampling diagnostic vs. thermodynamic temperature}

It is crucial to clarify what the configurational temperature measures in the context of Euclidean path integrals. 
Unlike in classical statistical mechanics, the configurational temperature does \emph{not} measure a thermodynamic temperature in the usual sense. 
Instead, it serves as a \emph{sampling consistency diagnostic} that verifies whether field configurations are generated with the correct statistical weight relative to the action, $e^{-{\cal S}[\phi]}$.

Introducing an explicit parameter, $\beta_{\rm config}$, in the statistical weight
\begin{equation}
P[\phi] \propto e^{-\beta_{\rm config} {\cal S}[\phi]},
\end{equation}
with the target value $\beta_{\rm config} = 1$ allows one to define a {\it configurational temperature estimator} directly in the field-theoretic setting.

If the sampling procedure is correct, the estimator reproduces the target value $\beta_{\rm config} = 1$.
Conversely, if the algorithm converges to an incorrect distribution, due to numerical instabilities, finite step-size effects, or failures of the complex Langevin process, one generically finds \cite{Dhindsa:2025xfv, Longia:2026doi}
\begin{equation}
\beta_{\rm config} \neq 1.
\label{eq:effective_coupling}
\end{equation}
The configurational temperature therefore provides a direct and quantitative probe of the correctness of the sampling.

The configurational temperature estimator can be applied directly to the lattice action.
Let us define the measured configurational temperature estimator in the lattice simulations as $\beta_M$. 
We have
\begin{equation}
\beta_M \equiv \Bigg\langle \vec{\nabla}_\phi \cdot \frac{\vec{\nabla}_\phi {\cal S}[\phi]}{|\vec{\nabla}_\phi {\cal S}[\phi]|^2} \Bigg\rangle,
\label{eq:lattice_estimator}
\end{equation}
where the gradient and Hessian are taken with respect to the lattice fields $\phi$ at each site, and the expectation value is computed over the ensemble of generated configurations.

The estimator in Eq.~\eqref{eq:lattice_estimator} contains the factor $|\vec{\nabla}_\phi {\cal S}[\phi]|^{-2}$, which could appear singular at stationary configurations where $\vec{\nabla}_\phi {\cal S}[\phi] = 0$. 
However, near such points the numerator vanishes with the same order. 
Expanding the action around a stationary configuration $\phi_0$,
\begin{equation}
{\cal S}(\phi) = {\cal S}(\phi_0) + \frac{1}{2}(\phi - \phi_0)^T H (\phi - \phi_0) + \cdots,
\end{equation}
one finds that the gradient scales linearly with the displacement $(\phi - \phi_0)$, while both numerator and denominator of the estimator scale quadratically. 
Their ratio therefore remains finite. In practice we observe no numerical instabilities associated with configurations close to stationary points of the action. 

Comparing $\beta_M$ with the expected value $\beta_{\mathrm{config}} = 1$ provides a stringent test of:
\begin{enumerate}
\item[$(i.)$] {\it Algorithmic correctness}: Whether the sampling dynamics correctly implements the target distribution,
\item[$(ii.)$] {\it Numerical stability}: Whether discretization errors (finite step size, finite lattice spacing) introduce systematic biases,
\item[$(iii.)$] {\it Thermalization}: Whether the system has reached equilibrium.
\end{enumerate}

\subsubsection{Finite-size and discretization effects}

The $\mathcal{O}(1/N)$ corrections indicated in Eq.~\eqref{eq:temp_eq} become relevant at finite lattice volumes.
For a lattice with $N_{\mathrm{sites}} = N_\tau \times N_s^{d-1}$ degrees of freedom (where $d$ is the spacetime dimension), ensemble equivalence holds only in the thermodynamic limit $N_{\mathrm{sites}} \to \infty$.
At finite volume, systematic deviations of order $1/N_{\mathrm{sites}}$ are expected and can be used to characterize lattice artifacts.

Additionally, discretization effects from finite lattice spacing $a$ can shift the effective action and modify the realized configurational temperature.
The comparison between $\beta_{\mathrm{config}}$ and $\beta_M$ therefore serves a dual purpose: ($i.$) it verifies sampling correctness (algorithmic validation) and ($ii.$) it quantifies discretization artifacts (lattice calibration).

This dual diagnostic capability is especially valuable in complex Langevin dynamics~\cite{Dhindsa:2025xfv, Joseph:2025xbn}, where subtle convergence failures can occur even when simulations appear stable by other metrics.
The power of this approach lies in its ability to detect sampling inconsistencies through a physically interpretable, computationally accessible observable that depends only on local properties of field configurations.

\section{One-dimensional PT-symmetric theories}
\label{sec:One-dimensional_PT-symmetric_theories}

It has long been observed~\cite{Bender:1998ke, Bender:1998gh} that quantum mechanical theories with PT-symmetric but non-Hermitian Hamiltonians nevertheless possess real, positive-definite spectra.  
A simple but nontrivial example is provided by the one-dimensional ($0+1$D) scalar field theory with potential
\begin{equation}
\label{eqn:lat-scalar-pt-symm-pot}
V(\phi) = - \frac{g}{N} \left( i \phi \right)^N,
\end{equation}
where the coupling $g > 0$ and $N = 2 + \delta$ with $\delta > 0$ following Ref.~\cite{Bender:1999ek}.

The Euclidean action is an integral over the thermal circle, with Euclidean time ($\tau$) compactified to circumference $\beta$:
\begin{equation}
{\cal S} = \int_0^\beta d\tau \, 
\left[ \tfrac{1}{2} \Big(\tfrac{\partial \phi}{\partial \tau}\Big)^2 + V(\phi) \right].
\end{equation}

This model exhibits features that are inaccessible to standard perturbative methods.
For instance, the one-point function, $G_1 \equiv \langle \phi \rangle$, in the PT-symmetric vacuum is nonzero and purely imaginary, a result that cannot be obtained from Feynman rules but can be derived systematically using Schwinger--Dyson truncations~\cite{Bender:1999ek}.  
The values thus obtained for the selected cases with $g = 1$ in Ref. \cite{Bender:1999ek} are
\begin{align}
G_1^{\rm SDE} &= - i \, 0.5901 \quad (\delta = 1), \\
G_1^{\rm SDE} &= - i \, 0.8669 \quad (\delta = 2),
\end{align}
while Ref. \cite{Pehlevan:2007eq}, using complex Langevin dynamics, gives
\begin{align}
G_1^{\rm GP} &= 0.0000(32) - i~0.8891(12) \quad (\delta = 2),
\end{align}
The full energy spectrum of the model remains real and bounded below.  
These features make the model a valuable nontrivial benchmark.

Because the potential is unbounded on the real axis, the path integral must be defined along a complex contour. 
Complex Langevin simulations have been shown to reproduce the corresponding contour solutions~\cite{Pehlevan:2007eq}.  
Multiple stationary distributions are possible depending on the initial conditions, but restricting initial fields to the lower half-plane selects the PT-symmetric solution~\cite{Bernard:2001wh}.

On a lattice with $N_\tau$ sites, the discretized action takes the form (we use the same notation, ${\cal S}$ for the continuum and lattice action)
\begin{equation}
{\cal S} = \sum_{n = 0}^{N_\tau - 1} \left[ \frac{(\phi_{n+1} - \phi_n)^2}{2} - \frac{g}{2 + \delta} \, (i \phi_n)^{2 + \delta} \right],
\end{equation}
with periodic boundary conditions for the scalar field.  
Denoting the lattice spacing as $a$, the dimensionless field is related to the physical one by $\phi = \phi_{\rm phys}/\sqrt{a}$, while the coupling scales as $g = a^{2 + \delta/2} g_{\rm phys}$.  
The circumference of the Euclidean time circle, which can be interpreted as inverse temperature (and not related to the configurational temperature $\beta_{\rm config}$), is 
\begin{equation}
\beta = N_\tau a. \nonumber
\end{equation}

The observables we are interested in are the equal-time correlation functions
\begin{equation}
G_k \equiv \langle \phi^k \rangle, \quad k = 1, \dots, 4,
\end{equation}
for $\delta = 1, 2$.  

\begin{figure*}[htbp]
\centering
\includegraphics[width=2.8in]{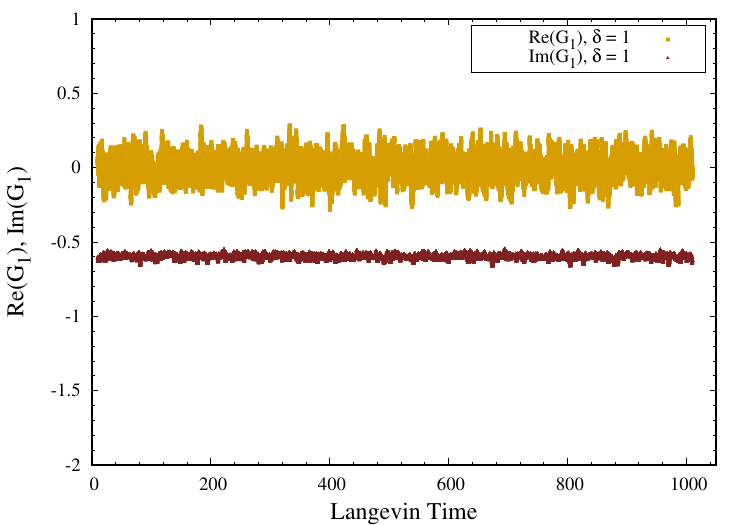}	
\includegraphics[width=2.8in]{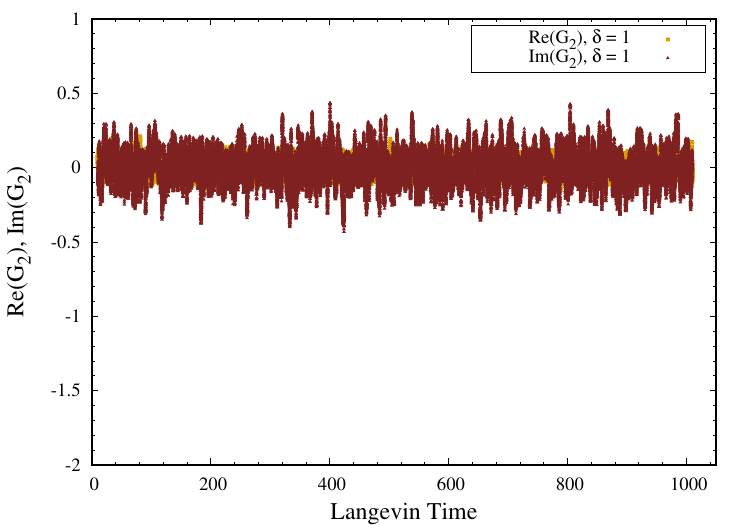}
\includegraphics[width=2.8in]{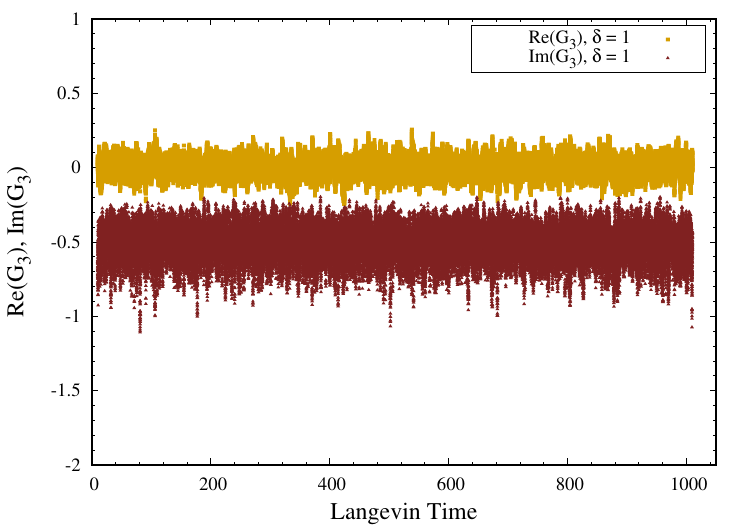}
\includegraphics[width=2.8in]{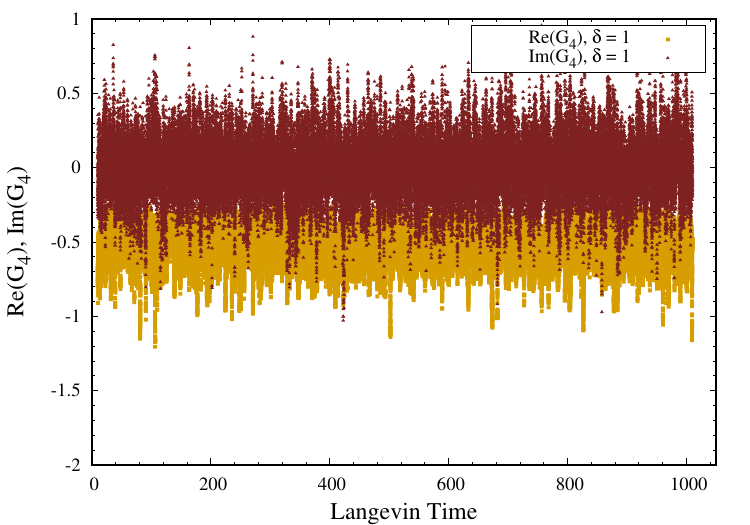}
\caption {Langevin time history of the observables $G_k$, $k = 1, 2, 3, 4$, for the PT-symmetric model with $\delta = 1$ and coupling $g = 1.0$.}
\label{fig:corr-bosonic-pt}
\end{figure*}

We considered lattices with fixed $N_\tau = 128$ and lattice spacing $a = \{ 0.0078$, $0.0156$, $0.0234$, $0.0313$, $0.0391$, $0.0469$, $0.0547$, $0.0625$, $0.0703$, $0.0781\}$. 
These lattice spacings correspond to the temporal extent $\beta = \{1, 2, \cdots, 10\}$.
We need to restrict the initial field to the lower half-plane to select the PT-symmetric solution~\cite{Bernard:2001wh}.
Thus, we started with $\phi_{\rm initial} = 0.0 - i 1.0$.
We used adaptive Langevin step size $\epsilon \equiv \Delta\theta \leq 0.001$, thermalization steps $N_{\rm therm} = 10^4$, and generation steps $N_{\rm gen} = 10^6$. 
Measurements were taken after every $10$ steps.
In Table~\ref{tab:lat-ptsymm-n3-n4} we show the results for the correlators.
They agree well with previous studies~\cite{Pehlevan:2007eq, Bernard:2001wh}, confirming that our CLM setup reliably reproduces the PT-symmetric stationary distribution.  
Figure~\ref{fig:corr-bosonic-pt} shows representative Langevin histories of the correlators.

\begin{table*}[htbp]
\centering
\begin{tabular}{c c	c}
\hline\hline 
\\
& $\delta = 1$ &  $\delta = 2$ \\ \\
\hline \hline
$G_1 \phantom{\Bigg|}\equiv \langle \phi \rangle$ & $-0.0021(30) - i 0.5994(07)$ &  $-0.0011(16) - i 0.8997(06)$ \\
$G_2 \phantom{\Bigg|}\equiv \langle \phi^2 \rangle$ & $0.0000(06) + i 0.0033(43)$ &  $-0.5545(24) + i 0.0024(33)$ \\
$G_3 \phantom{\Bigg|}\equiv \langle \phi^3 \rangle$ & $0.0014(20) - i 0.5157(21)$ &  $0.0034(42) + i 0.0000(11)$ \\
$G_4 \phantom{\Bigg|}\equiv \langle \phi^4 \rangle$ & $-0.5162(31) - i 0.0043(43)$ &  $-0.6330(24) - i 0.0030(32)$ \\
\hline
\end{tabular}
\caption{The correlation functions $G_k$, $k = 1, 2, 3, 4$, for the PT-symmetric model with $\delta = 1, 2$ and coupling $g = 1.0$.}
\label{tab:lat-ptsymm-n3-n4}
\end{table*}

\subsection{Configurational temperature estimator}
\label{sec:Temperature_estimator}

We now evaluate the configurational temperature estimator in this PT-symmetric model.  
We denote the estimator for the $i$-th configuration in the Langevin evolution steps as $\hat{\beta}_i$.
Then, we have
\begin{equation}
\hat{\beta}_i \equiv
\frac{\sum_n h_{nn}}{\sum_n g_n^2} 
- \frac{2 \sum_{nm} g_n g_m h_{nm}}{\left(\sum_n g_n^2\right)^2}.
\end{equation}
It is the lattice discretization of Eq.~\eqref{eq:hessian-form}, with the elements of the gradient and Hessian defined as
\begin{equation}
g_n \equiv \frac{\partial {\cal S}}{\partial \phi_n}, 
\qquad 
h_{nm} \equiv \frac{\partial^2 {\cal S}}{\partial \phi_n \partial \phi_m}.
\end{equation}
Explicitly,
\begin{align}
g_n &= \Big[ (\phi_n - \phi_{n-1}) + (\phi_n - \phi_{n+1}) - g\, i^{2+\delta} \phi_n^{1+\delta}\Big], \\
h_{nn} &= \Big[ 2 - g (1+\delta) i^{2+\delta} \phi_n^{\delta} \Big], \\
h_{nm} &= - \big[ \delta_{m,n-1} + \delta_{m,n+1} \big],
\end{align}
where $\delta_{m,n}$ denotes the Kronecker delta, and periodic boundary conditions along the $\tau$ direction are imposed.

The measured configurational temperature is then defined as the real part of $\hat{\beta}_i$, averaged over the configurations:
\begin{equation}
\beta_M \equiv \mathrm{Re}\left[ \frac{1}{N_{\mathrm{config}}} \sum_{i=1}^{N_{\mathrm{config}}} \hat{\beta}_i \right].
\end{equation}

\begin{figure*}[htbp]
\centering
\includegraphics[width=3.0in]{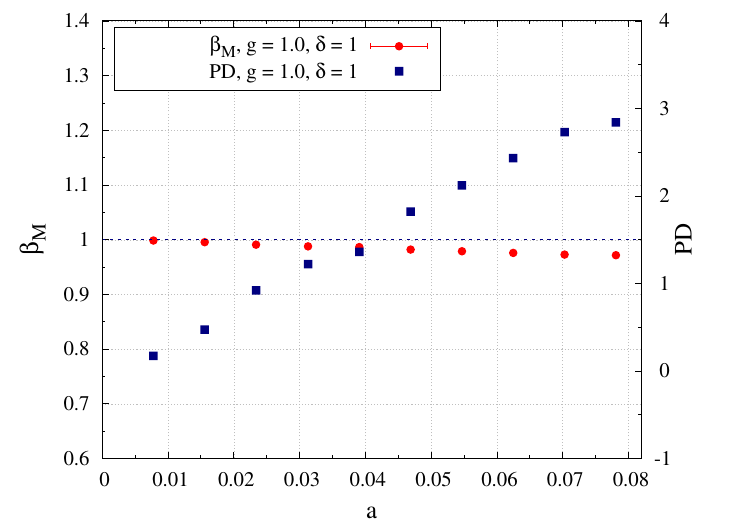}	
\includegraphics[width=3.0in]{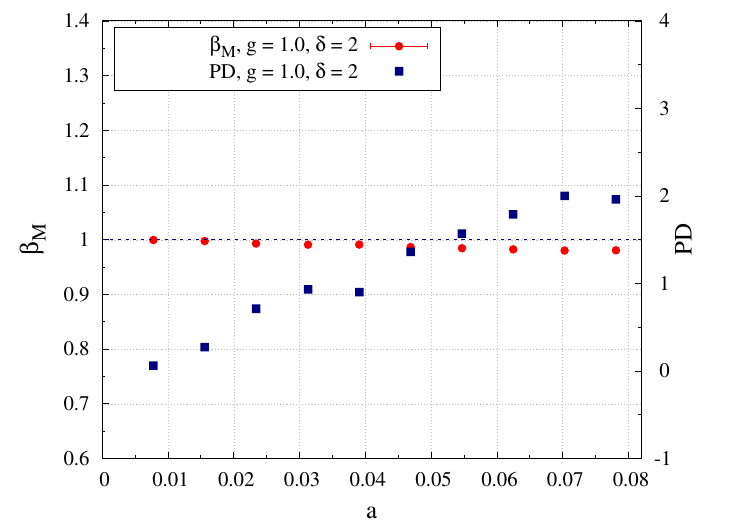}
\caption {The estimated configurational temperature $\beta_M$ and the percentage deviation in estimated temperature, ${\rm PD} \equiv 100 \times (\beta_{\rm config} - \beta_M)/\beta_{\rm config}$ against lattice spacing $a$. The data are for the PT-symmetric model given in Eq. \eqref{eqn:lat-scalar-pt-symm-pot}. The systematic growth of the relative error with $a$ reflects finite lattice spacing and discretization effects, demonstrating the estimator's sensitivity to lattice artifacts.}
\label{fig:beta_beta_m_delta}
\end{figure*}

Table~\ref{tab:lat-ptsymm-beta-n3-n4} shows $\beta_M$ compared with the target value $\beta_{\rm config} = 1$ for $\delta = 1, 2$.
The imaginary part of $\hat{\beta}_i$ averaged over the configurations is consistent with zero and has been omitted for clarity. 
Figure~\ref{fig:beta_beta_m_delta} displays the dependence of $\beta_M$ on $\beta_{\rm config}$ and the relative error $(\beta_{\rm config} - \beta_M) / \beta_{\rm config}$, demonstrating the accuracy and stability of the estimator across a range of lattice spacings.

The estimator on the lattice, $\beta_M$, reproduces the expected value $\beta_{\rm config} = 1$ within 0.2 - 3\% across the full range tested. 
The systematic deviation grows with the lattice spacing $a$, reaching approximately 2.8\% at $a \simeq 0.08$.
They can arise due to finite Langevin step-size and finite lattice spacing/discretization effects. 

\begin{table*}[htbp]
\centering
\begin{tabular}{c c c c c c}
\hline\hline 
\\
$~~~ \beta_{\rm config} ~~~$ & $~~~ a ~~~$ & $\beta_M (\delta = 1)$  & PD (\%) & $\beta_M (\delta = 2)$ & PD (\%) \\ \\
\hline \hline \\
$1.00$ & $0.0078$ & $0.9983(14)$  & $0.17$ & $0.9994(22)$ & $0.06$ \\  
\\
$1.00$ & $0.0156$ &  $0.9954(15)$  & $0.47$ & $0.9974(22)$ & $0.27$ \\  
\\
$1.00$ & $0.0234$ &  $0.9908(05)$  & $0.92$ & $0.9929(07)$ & $0.71$ \\  
\\
$1.00$ & $0.0313$ &  $0.9878(06)$  & $1.22$ & $0.9907(09)$ & $0.93$ \\  
\\
$1.00$ & $0.0391$ &  $0.9864(15)$  & $1.36$ & $0.9910(22)$ & $0.90$ \\  
\\
$1.00$ & $0.0469$ &  $0.9818(05)$  & $1.82$ & $0.9864(07)$ & $1.36$ \\  
\\
$1.00$ & $0.0547$ &  $0.9788(05)$  & $2.12$ & $0.9843(07)$ & $1.57$ \\  
\\
$1.00$ & $0.0625$ &  $0.9758(05)$  & $2.43$ & $0.9821(07)$ & $1.79$ \\  
\\
$1.00$ & $0.0703$ &  $0.9727(05)$  & $2.73$ & $0.9798(07)$ & $2.00$ \\  
\\
$1.00$ & $0.0781$ &  $~~0.9716(14)~~$  & $2.84$ & $~~0.9804(22)~~$ & $1.96$ \\  
\\
\hline
\end{tabular}
\caption{The expected temperature $\beta_{\rm config}$ and the estimated temperature $\beta_M$ for the model given in Eq.~\eqref{eqn:lat-scalar-pt-symm-pot} for $\delta = 1, 2$ and coupling $g = 1.0$. The percentage deviation ${\rm PD} \equiv 100 \times (\beta_{\rm config} - \beta_M)/\beta_{\rm config}$ quantifies the systematic underestimation, which grows with lattice spacing $a$ and reflects finite lattice spacing and discretization effects as discussed in Sec.~\ref{sec:Application_to_Euclidean_lattice_field_theory}. See also Fig.~\ref{fig:beta_beta_m_delta}.}
\label{tab:lat-ptsymm-beta-n3-n4}
\end{table*}

\section{Numerical Tests of the Estimator}
\label{sec:Numerical_Tests_of_the_Estimator}

One of the most appealing features of the configurational temperature estimator is its direct sensitivity to algorithmic errors and numerical artifacts.  
Since it probes sampling consistency, any deviation of the measured configurational temperature $\beta_M$ from the target $\beta_{\rm config}$ provides an immediate signal of potential problems.  
In this section, we illustrate its utility in three contexts: detecting algorithmic errors, assessing Langevin step-size dependence, and monitoring thermalization.

\subsection{Detecting algorithmic errors}
\label{sec:Detecting_algorithmic_errors}

As a first test, we deliberately introduce an error into the noise normalization of the discretized Langevin equation.  
Specifically, we replace the correct Gaussian correlation given in Eq. \eqref{eq:Gaussian_noise}
\begin{equation}
\langle \eta(\theta) \eta(\theta') \rangle = 2 \, \delta_{\theta \theta'} \nonumber
\end{equation}
with
\begin{equation}
\langle \eta(\theta) \eta(\theta') \rangle = \sigma \, \delta_{\theta \theta'},
\end{equation}
where $\sigma > 0, \sigma \neq 2$, represents a mis-scaling of the noise term. 
In stochastic quantization, the mis-scaling, analogous to the rescaling of the diffusion coefficient in Brownian motion~\cite{Uhlenbeck:1930zz, Kubo:1966fyg}, breaks the fluctuation-dissipation balance, thereby modifying the stationary distribution of the associated Fokker--Planck equation from $e^{-{\cal S}}$ to $e^{-(2 / \sigma) {\cal S}}$~\cite{Damgaard:1987rr}. 
In this setup, the white-noise normalization $\sigma$ effectively sets the thermostat; departures from the canonical value $\sigma = 2$ imply that the dynamics samples at an effective configurational temperature 
\begin{equation}
\beta_{\rm config}^{\rm eff} = (2 / \sigma)~\beta_{\rm config}, \nonumber
\end{equation}
which our configuration-based thermometer is designed to register.

As shown in Table~\ref{tab:lat_ptsymm_beta_algo_err}, this modification is clearly detected by the configurational estimator: $\beta_M$ shifts systematically away from the input value $\beta_{\rm config}$ in a manner proportional to $\sigma$.  
Even modest changes, such as $\sigma=1.0$, produce a measurable deviation.  
Thus, the estimator serves as a robust diagnostic for errors that might otherwise remain hidden.

The predicted scaling $\beta_M = (2/\sigma) \beta_{\rm config}$ is verified numerically to high precision. 
For example, at $\sigma = 0.5$, we obtain $\beta_M = 3.9888(59)$, compared to the predicted value $(2.0 / 0.5) \times 1.0 = 4.0$, representing $(4.0 - 3.9888) / 4.00 \approx 0.3\%$ agreement. 
This quantitative confirmation validates Eq. \eqref{eq:effective_coupling} and demonstrates the effectiveness of the estimator as a sampling diagnostic.

\begin{table*}[htbp]
\centering
\begin{tabular}{c c c}
\hline\hline
\\
$~~~ \sigma ~~~$ & $\beta_M (\delta = 1)$  & $\beta_M (\delta = 2)$ \\ \\
\hline \hline \\
$0.5$ & $3.9888(59)$  & $3.9941(64)$ \\ \\
$1.0$ & $1.9953(29)$  & $1.9980(36)$ \\ \\
$2.0$ & $0.9983(14)$  & $0.9994(22)$ \\ \\
$4.0$ & $0.4996(7)$   & $0.4986(14)$ \\ \\
\hline
\end{tabular}
\caption{Deviation of the measured configurational temperature $\beta_M$ as $\sigma$ is changed from its standard value $\sigma = 2.0$. As predicted by Eq.~\eqref{eq:effective_coupling}, the noise mis-scaling modifies the sampled distribution from $e^{-{\cal S}}$ to $e^{-(2/\sigma){\cal S}}$, resulting in $\beta_M = (2/\sigma)\beta_{\rm config}$. The configurational temperature estimator thus correctly detects and quantifies this algorithmic error.}
\label{tab:lat_ptsymm_beta_algo_err}
\end{table*}

\subsection{Dependence on Langevin step size}
\label{sec:Dependence_on_Langevin_step_size}

We next examine the effect of the Langevin step size $\epsilon$.  
In principle, correct results should be recovered in the $\epsilon \to 0$ limit, but finite step sizes introduce discretization artifacts.  
We show in Table~\ref{tab:fixed_step_size} that $\beta_M$ deviates increasingly from the target value as $\epsilon$ grows, with convergence eventually breaking down (e.g., for $\delta=2$ at $\epsilon \gtrsim 0.1$). 
The sensitivity of $\beta_M$ to $\epsilon$ provides a clear and quantitative criterion for selecting a step size small enough to suppress discretization errors, complementing standard checks based on observables alone.

\begin{table*}[htbp]
\centering
\begin{tabular}{c c c c}
\hline\hline
\\
$~~~ \beta_{\rm config} ~~~$ & $~~~ \epsilon ~~~$ & $\beta_M (\delta = 1)$  & $\beta_M (\delta = 2)$ \\ \\
\hline \hline \\
\multirow{11}{*}{$1.00$} 
 & $0.0001$ & $1.0002(15)$  & $0.9845(22)$ \\ \\
 & $0.001$ & $0.9983(14)$  & $0.9994(22)$ \\ \\
 & $0.01$ & $0.9830(15)$  & $0.9898(23)$ \\ \\
 & $0.03$ & $0.9474(14)$  & $0.9663(29)$ \\ \\
 & $0.1$ & $0.8251(12)$   & $-$ \\ \\
 & $0.2$ & $0.6437(11)$    & $-$ \\ \\
\hline
\end{tabular}
\caption{Deviation of the measured configurational temperature $\beta_M$ for various values of the fixed Langevin step size $\epsilon$. The configurational temperature estimator shows clear sensitivity to discretization errors: $\beta_M$ deviates increasingly from the target value as $\epsilon$ grows, with convergence eventually breaking down for $\delta = 2$ at $\epsilon \gtrsim 0.1$ (indicated by hyphens). This demonstrates that the estimator provides a quantitative criterion for selecting step sizes small enough to suppress discretization artifacts.}
\label{tab:fixed_step_size}
\end{table*}

\subsection{Monitoring thermalization}
\label{sec:Monitoring_thermalization}

We test the ability of the estimator to monitor thermalization by examining its behavior during the initial stages of Langevin evolution. 
As the system equilibrates, both physical observables and $\beta_M$ gradually approach their equilibrium values from the initial PT-symmetric configuration, $\phi_{\mathrm{initial}} = 0.0 - i1.0$. 

\begin{figure*}[htbp]
\centering
\includegraphics[width=3.0in]{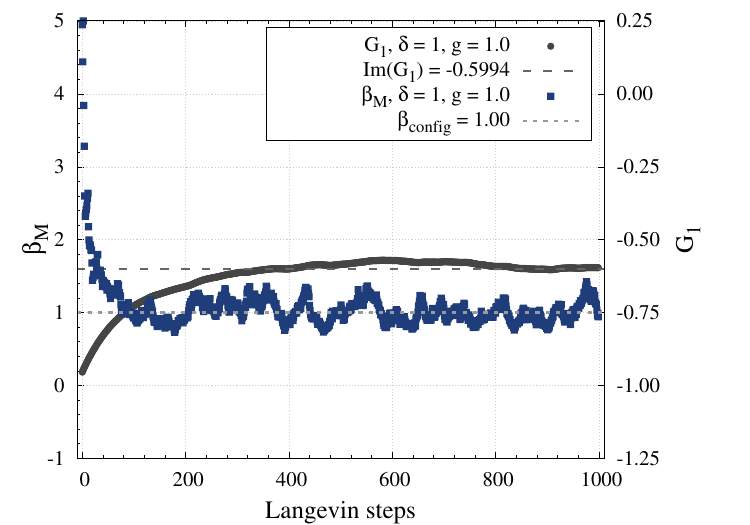}	
\includegraphics[width=3.0in]{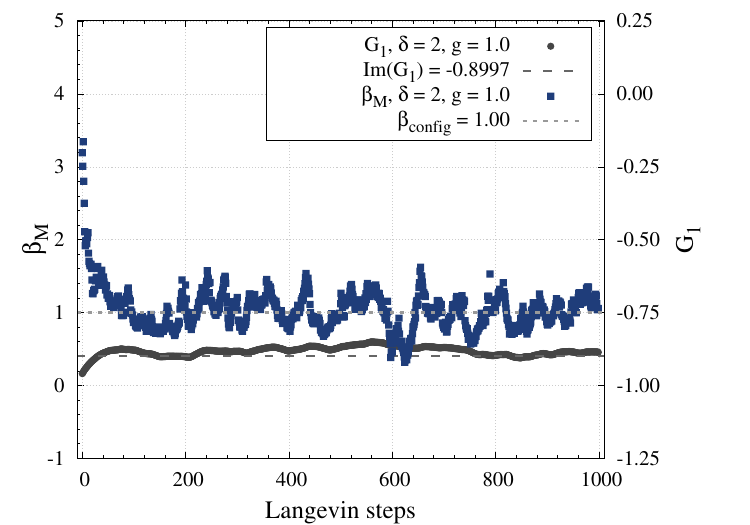}
\caption{Thermalization behavior of the one-point function $G_1 \equiv \langle\phi\rangle$ (imaginary part) and the measured configurational temperature estimator $\beta_M$ during the initial Langevin evolution. The data are for the PT-symmetric model with $\delta = 1$ (left) and $\delta = 2$ (right) at coupling $g = 1.0$.}
\label{fig:therm_behaviour}
\end{figure*}

Figure~\ref{fig:therm_behaviour} shows this behavior for the one-point function, $G_1 \equiv \langle\phi\rangle$, alongside $\beta_M$ for both $\delta = 1$ and $\delta = 2$. 
The configurational temperature estimator exhibits thermalization dynamics that closely track those of the physical observable, with both quantities settling into stable distributions within approximately 300-400 Langevin steps. 
Notably, $\beta_M$ converges to the correct value $\beta_{\rm config} = 1.00$ (within statistical fluctuations) on the same timescale as $G_1$ reaches its equilibrium value.

This parallel thermalization behavior has important practical implications. 
Since $\beta_M$ can be computed at every Langevin step, it provides a valuable real-time diagnostic for assessing equilibration. 
Unlike physical observables, whose equilibrium values may be unknown a priori in unexplored theories, $\beta_M$ has a known target value, making it particularly useful for verifying that sufficient thermalization has been achieved before measurements begin. 
The estimator thus serves as both a correctness check and a practical guide for determining appropriate thermalization times.

\section{Comparison with Existing Diagnostics}
\label{sec:Comparison_with_Existing_Diagnostics}

Different correctness criteria for complex Langevin simulations probe distinct aspects of the method. 
Diagnostics such as the decay of the drift distribution or the analysis of boundary terms test the validity of the analytic continuation connecting the complex Langevin process on the complexified manifold to the original path integral with complex measure. 

By contrast, the configurational temperature estimator introduced in this work probes the internal consistency of the stochastic sampling process itself. 
In particular, it tests whether the dynamics samples configurations with the intended statistical weight $e^{-{\cal S}}$. 

These diagnostics therefore address complementary questions: the configurational temperature tests the correctness of the stochastic sampling procedure, while drift-based and boundary-term criteria probe whether the resulting distribution corresponds to the correct path integral after analytic continuation.

\subsection{Langevin operator on observables}
\label{sec:Langevin_operator_on_observables}

A correctness criterion for the CLM is based on the Langevin operator acting on observables~\cite{Aarts:2011ax, Aarts:2009uq, Aarts:2013uza}.  
For an observable ${\cal O}_n[\phi,\theta]$ at lattice site $n$, we have
\begin{equation}
\frac{\partial {\cal O}_n [\phi, \theta]}{\partial \theta} = L_n \, {\cal O}_n[\phi, \theta], 
\end{equation}
where the operator $L_n$ is defined as
\begin{equation}
L_n \equiv \left( \frac{\partial}{\partial \phi_n} - \frac{\partial S}{\partial \phi_n} \right) \frac{\partial}{\partial \phi_n}.
\end{equation}

In equilibrium, the integration by parts argument implies that
\begin{equation}
C_{{\cal O}_n} \equiv \big\langle L_n {\cal O}_n[\phi] \big\rangle = 0,
\end{equation}
which can be used as a diagnostic of correctness.  

\begin{figure*}[htbp]
\centering
\includegraphics[width=3.0in]{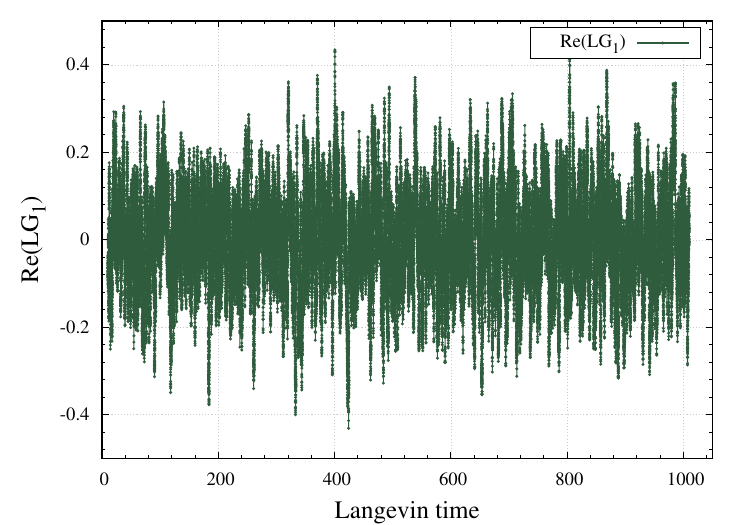}	
\includegraphics[width=3.0in]{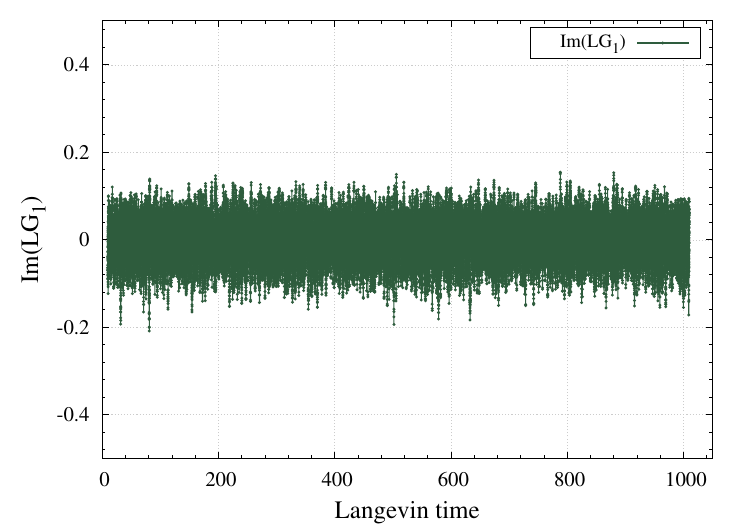}
\caption {The Langevin time history of $L G_1$. The data are for the PT-symmetric model with $\delta = 1$ and coupling $g = 1.0$.}
\label{fig:LG1_Lang_history}
\end{figure*}

We apply this criterion to the one-point function $G_1 \equiv \langle \phi \rangle$.  
Due to translational symmetry, $L G_1$ can be averaged over all sites.  
Figure~\ref{fig:LG1_Lang_history} shows the Langevin-time history of $L G_1$, and Table~\ref{tab:comparison_three_criteria} summarizes the results.  
Within errors, $L G_1$ is consistent with zero, suggesting that this criterion does not flag problems in the simulations considered here.

\subsection{Decay of the drift terms}
\label{sec:Decay_of_the_drift_terms}

A second criterion, introduced by Nagata {\it et al.}~\cite{Nagata:2016vkn, Nagata:2018net}, requires the probability distribution of the drift magnitude to decay exponentially or faster at large values.  
This ensures that integration by parts is valid and convergence is correct.

\begin{figure*}[htbp]
\centering
\includegraphics[width=3.0in]{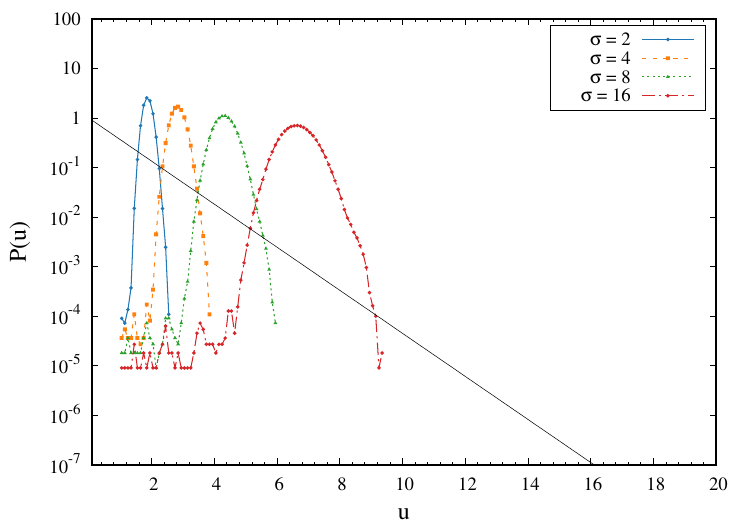}	
\includegraphics[width=3.0in]{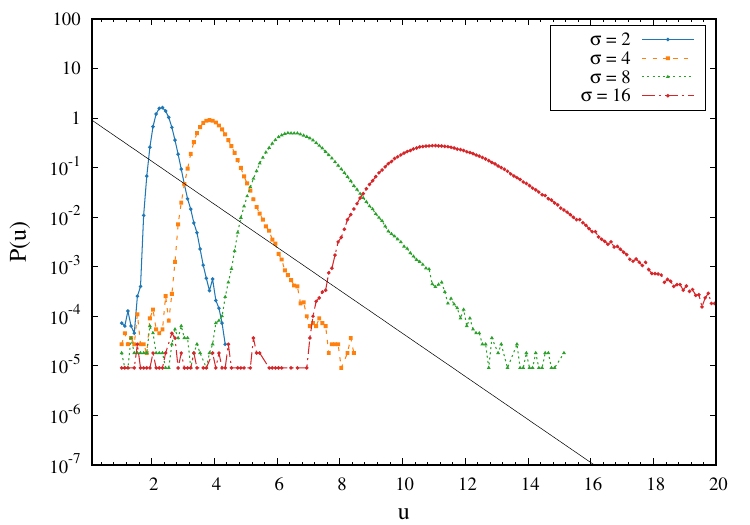}
\caption {Decay of the drift terms. The simulations are for the PT-symmetric model with $\delta = 1$ (left) and $\delta = 2$ (right) and coupling $g = 1.0$.}
\label{fig:Pu_u_g1_d1_d2_s_2_4_8_16}
\end{figure*}

The magnitude of the mean drift is defined as
\begin{equation}
u \equiv \sqrt{\frac{1}{N_\tau} \sum_{n=0}^{N_\tau-1} 
\left| \frac{\partial S}{\partial \phi_n} \right|^2 }.
\end{equation}
Figure~\ref{fig:Pu_u_g1_d1_d2_s_2_4_8_16} shows the drift distribution $P(u)$ for several noise variances $\sigma$.  
For $\delta=1$, exponential suppression holds for all tested $\sigma$.  
For $\delta=2$, exponential decay is observed at $\sigma=2,4,8$, but breaks down at $\sigma=16$, indicating unreliable dynamics in that case.

\subsection{Comparing the three correctness criteria}
\label{sec:Comparing_the_three_correctness_criteria}

To assess whether each criterion is satisfied, we adopt the following 
quantitative thresholds:
\begin{itemize}
\item[$(i.)$] {\it Criterion 1}: Satisfied if $| \beta_M - \beta_{\rm config}| < 3 \sigma_{\rm stat}$, where $\sigma_{\rm stat}$ is the statistical uncertainty in $\beta_M$,
\item[$(ii.)$] {\it Criterion 2}: Satisfied if $|\langle LG_1 \rangle| < 3 \sigma_{\rm stat}$,
\item[$(iii.)$] {\it Criterion 3}: Satisfied if $P(u)$ exhibits exponential or faster decay, as determined by visual inspection and consistency with exponential fits.
\end{itemize}

With $\sigma = 2.0$ (correct noise normalization), all three criteria are satisfied. 
For $\sigma \neq 2$, only Criterion 1 reliably detects the algorithmic error.

We probe these criteria by introducing algorithmic errors via controlled noise mis-scaling.  
The results are summarized in Table~\ref{tab:comparison_three_criteria}.  
We find that:
\begin{itemize}
\item[$(i.)$] {\it Criterion 1} is highly sensitive: deviations in noise normalization are immediately reflected in $\beta_M \neq \beta_{\rm config}$.  
\item[$(ii.)$] {\it Criterion 2} is satisfied in all cases, but its insensitivity means it fails to flag algorithmic errors.  
\item[$(iii.)$] {\it Criterion 3} is generally satisfied, but breaks down only at extreme mis-scalings (e.g. $\sigma=16$ for $\delta=2$).  
\end{itemize}

Thus, the configurational temperature emerges as a sharper diagnostic, capable of detecting subtle errors that remain invisible to existing criteria.  
Its interpretation as a sampling consistency check further distinguishes it as an independent and complementary test of CLM reliability. 
The configurational temperature estimator can identify noise mis-scaling, expose step-size artifacts, and track thermalization, providing information complementary to existing drift-based diagnostics while remaining easy to implement in practice.

\begin{table*}[htbp]
\centering
\begin{tabular}{c c c c c c}
\hline\hline
\\
 &  &  & $~$ Criterion 1 & Criterion 2 & Criterion 3 \\ 
$~ \delta ~$ & $~ \beta_{\rm config} ~$ & $\sigma$ & $\beta_M = \beta_{\rm config}$ & $\langle L G_1 \rangle = 0$ & Drift term fall-off \\ \\
\hline \hline \\
\multirow{10}{*}{$1.0$} & \multirow{10}{*}{$1.0$}
  & $2.0$  & \makecell{$0.9983(14)$ \\ (Satisfied)} 
            & \makecell{$-0.0010(3)$ \\ (Satisfied)} & \makecell{Exponential \\ (Satisfied)} \\  \\
 & & $4.0$  & \makecell{$0.4984(2)$ \\ (Not Satisfied)} 
            & \makecell{$-0.0014(5)$ \\ (Satisfied)} & \makecell{Exponential \\ (Satisfied)} \\  \\
 & & $8.0$  & \makecell{$0.2492(1)$ \\ (Not Satisfied)} 
            & \makecell{$-0.0019(9)$ \\ (Satisfied)} & \makecell{Exponential \\ (Satisfied)} \\  \\
 & & $16.0$ & \makecell{$0.1246(0)$ \\ (Not Satisfied)} 
            & \makecell{$-0.0027(14)$ \\ (Satisfied)} & \makecell{Exponential \\ (Satisfied)} \\
\\ \hline \\
\multirow{10}{*}{$2.0$} & \multirow{10}{*}{$1.0$}
  & $2.0$  & \makecell{$0.9994(22)$ \\ (Satisfied)} 
            & \makecell{$-0.0010(5)$ \\ (Satisfied)} & \makecell{Exponential \\ (Satisfied)} \\  \\
 & & $4.0$  & \makecell{$0.4975(12)$ \\ (Not Satisfied)} 
            & \makecell{$-0.0013(8)$ \\ (Satisfied)} & \makecell{Exponential \\ (Satisfied)} \\  \\
 & & $8.0$  & \makecell{$0.2485(3)$ \\ (Not Satisfied)} 
            & \makecell{$-0.0018(15)$ \\ (Satisfied)} & \makecell{Exponential \\ (Satisfied)} \\  \\
 & & $16.0$ & \makecell{$0.1238(2)$ \\ (Not Satisfied)} 
            & \makecell{$-0.0023(25)$ \\ (Satisfied)} & \makecell{Non-exponential \\ (Not Satisfied)} \\  \\
\\
\hline
\end{tabular}
\caption{Comparison of the three complementary correctness criteria in complex Langevin simulations for the PT-symmetric model with $\delta = 1, 2$ and coupling $g = 1.0$.}
\label{tab:comparison_three_criteria}
\end{table*}

Another important diagnostic is the study of boundary terms arising in the formal justification of the complex Langevin method. 
When the probability distribution on the complexified manifold does not decay sufficiently fast at large imaginary field values, integration by parts in the derivation of the Schwinger--Dyson relations generates boundary contributions. 
These boundary terms signal a breakdown of the formal equivalence between the complex Langevin process and the original path integral. 
Methods for monitoring such boundary terms have been developed in Refs. \cite{Aarts:2009uq, Aarts:2011ax, Scherzer:2018hid, Scherzer:2019lrh}.

\subsection{Computational cost and scalability}

The configurational temperature estimator requires computing the gradient at each measured configuration. 
For a lattice with $N_{\rm sites}$ degrees of freedom the gradient costs ${\cal O}(N_{\rm sites})$ operations (computed routinely in CLM), the full Hessian costs ${\cal O}(N^2_{\rm sites})$ storage and computation, and the estimator evaluation costs ${\cal O}(N^2_{\rm sites})$ for the directional term.

In our 1D simulations ($N_\tau = 128$), the overhead is negligible (< 1\% of total simulation time). 
However, for 4D lattice gauge theory with $N_{\rm sites} \sim 10^6$, storing and computing the full Hessian becomes impractical.

Nevertheless, we can resort to scalable approximations.
Several strategies can reduce computational cost in large systems:

\textit{(i) Trace-only approximation}: In high dimensions, the second term in Eq. \eqref{eq:hessian-form} becomes subdominant. 
Using only $\mathrm{Tr}(\mathbb{H}) / |\vec{g}|^2$ reduces cost to ${\cal O}(N_{\rm sites})$.

\textit{(ii) Sparse Hessian structure}: For local actions, $\mathbb{H}$ is sparse (banded or block-diagonal), enabling efficient sparse matrix operations.

\textit{(iii) Stochastic trace estimation}: The trace can be estimated stochastically using random vectors via $\mathrm{Tr}(\mathbb{H}) \approx \sum_i v_i^T \mathbb{H} v_i$ for Rademacher vectors $v_i$, reducing cost to ${\cal O}(N_{\rm sites})$ per sample.

These approximations will be explored in future work on higher-dimensional theories.

\section{Discussion and Outlook}
\label{sec:Discussion_and_Outlook}

In this work, we introduced a configurational temperature estimator as a reliability diagnostic for the complex Langevin method. 
Our results in one-dimensional PT-symmetric models demonstrate that this estimator faithfully reproduces the input configurational temperature ($\beta_{\rm config} = 1$) to within 0.2 - 3\% (Table~\ref{tab:lat-ptsymm-beta-n3-n4}), 
sensitively detects algorithmic errors (Table~\ref{tab:lat_ptsymm_beta_algo_err}), and complements existing drift-based diagnostics. 
Crucially, the direct comparison in Table~\ref{tab:comparison_three_criteria} reveals that the configurational estimator appears to be more sensitive than either the Langevin operator  criterion or drift decay analysis in detecting noise mis-scaling errors. 
It provides an independent, physically interpretable cross-check of CLM dynamics by directly probing whether configurations are sampled with the correct statistical weight $e^{-{\cal S}[\phi]}$.

The choice of one-dimensional PT-symmetric systems as our test case is deliberate. 
These models serve as controlled benchmarks where correctness criteria can be evaluated against well-established results, allowing us to isolate algorithmic effects without the complications of higher-dimensional gauge theories. 
However, the method is not inherently limited to low dimensions. 
The estimator's local nature suggests that natural approximation strategies can be applied for larger systems.

A practical concern for extending this diagnostic to higher-dimensional theories is computational cost. 
Computing the full Hessian scales as $\mathcal{O}(N_{\mathrm{sites}}^2)$ in storage and operations, which becomes prohibitive for 4D gauge theories with $N_{\mathrm{sites}} \sim 10^6$. However, several mitigation strategies are available.
They are $(i.)$ {\it Trace-only approximation:} In high dimensions, the directional term in Eq.~\eqref{eq:hessian-form} becomes subdominant, allowing use of only $\mathrm{Tr}(\mathbb{H})/|\vec{g}|^2$, which reduces cost to $\mathcal{O}(N_{\mathrm{sites}})$; $(ii.)$ {\it Sparse structure:} Local lattice actions produce sparse Hessians  (banded or block-diagonal), enabling efficient sparse matrix methods; and $(iii.)$ {\it Stochastic trace estimation:} The trace can be estimated using random vectors: $\mathrm{Tr}(\mathbb{H}) \approx \sum_i \vec{v}_i^T \mathbb{H} \vec{v}_i$ for Rademacher vectors $\vec{v}_i$, reducing cost to $\mathcal{O}(N_{\mathrm{sites}})$ per sample.

Preliminary tests in 3D gauge theories~\cite{Joseph:2025xbn} suggest that these approximations maintain diagnostic sensitivity while remaining computationally tractable. 
Detailed benchmarking in 4D QCD remains an important direction for future work.

The configurational temperature diagnostic could be integrated into a hybrid reliability framework for CLM, complementing existing tools.

The immediate avenues for explorations are the following.
$(i.)$ {\it Lattice QCD at finite density}: Where the sign problem is most severe and existing diagnostics may be insufficient, $\beta_M$ provides an independent sampling consistency check. 
Deviations from $\beta_{\rm config} = 1$ would signal either algorithmic failure or unexpected effective action modifications; $(ii.)$ {\it Matrix models and supersymmetric theories:} Where modest system sizes keep computational overhead low, the estimator offers a real-time monitor during explorations of phase structure and spontaneous symmetry breaking; $(iii.)$ {\it Phase transition detection:} Sharp changes in $\beta_M$ at phase boundaries could provide early warning of sampling difficulties or serve as complementary indicators of thermal transitions; and $(iv.)$ {\it Finite-temperature field theories:} The consistency $\beta_M = \beta_{\rm config}$ within $\mathcal{O}(1/N_{\mathrm{sites}})$ corrections provides a stringent test. 
Deviations beyond expected finite-size  effects would indicate algorithmic instabilities or discretization artifacts.

The key advantage of configurational temperature over existing CLM diagnostics is its \emph{direct connection to sampling consistency}. 
While drift-based criteria probe kinetic properties of the Langevin evolution, $\beta_M$ directly tests whether the sampled ensemble has the correct statistical weights. 
This conceptual distinction makes it a natural complement to existing tools, as demonstrated by the comparative analysis in Sec.~\ref{sec:Comparing_the_three_correctness_criteria}.

Moreover, because $\beta_M$ has a known target value ($\beta_{\rm config} = 1$), it provides a universal diagnostic applicable to any theory, even where the equilibrium values of the physical observables are unknown. 
This universality, combined with its sensitivity to subtle algorithmic errors, positions configurational temperature as a valuable addition to the CLM reliability toolkit.

We view this work as the first step toward integrating configurational diagnostics into complex-action simulations. 
The ultimate goal is enabling reliable first-principles explorations of QCD at finite density, where robust diagnostics are essential for ensuring that numerical results reflect true physics rather than algorithmic artifacts.

Future work should focus on: $(i.)$ systematic benchmarking in higher-dimensional scalar and gauge theories; $(ii.)$ quantifying the trade-offs between computational cost and diagnostic sensitivity for various approximation schemes; $(iii.)$ exploring whether $\beta_M$ can serve as an order parameter or early-warning signal for phase transitions in complex-action theories. 
These developments will determine the practical scope of configurational monitoring in lattice field theory.

\acknowledgments
We extend our gratitude to Navdeep Singh Dhindsa, Piyush Kumar, and Vamika Longia for their invaluable discussions. 
The work of A.J. was supported in part by a Start-up Research Grant from the University of the Witwatersrand. 
A.J. gratefully acknowledges the warm hospitality of the National Institute for Theoretical and Computational Sciences (NITheCS) and Stellenbosch University during the NITheCS Focus Area Workshop, {\it Decoding the Universe: Quantum Gravity and Quantum Fields.} 
The work of A.K. was partly supported by the National Natural Science Foundation of China under Grants No. 12293064, No. 12293060, and No. 12325508, as well as the National Key Research and Development Program of China under Contract No. 2022YFA1604900.

\raggedright
\bibliographystyle{utphys}
\bibliography{bibfile}

@article{Ito:2020mys,
    author = "Ito, Yuta and Matsufuru, Hideo and Namekawa, Yusuke and Nishimura, Jun and Shimasaki, Shinji and Tsuchiya, Asato and Tsutsui, Shoichiro",
    title = "{Complex Langevin calculations in QCD at finite density}",
    eprint = "2007.08778",
    archivePrefix = "arXiv",
    primaryClass = "hep-lat",
    reportNumber = "KEK-TH-2230, RIKEN-QHP-479",
    doi = "10.1007/JHEP10(2020)144",
    journal = "JHEP",
    volume = "10",
    pages = "144",
    year = "2020"
}

@article{Sexty:2013ica,
    author = "Sexty, D{\'e}nes",
    title = "{Simulating full QCD at nonzero density using the complex Langevin equation}",
    eprint = "1307.7748",
    archivePrefix = "arXiv",
    primaryClass = "hep-lat",
    doi = "10.1016/j.physletb.2014.01.019",
    journal = "Phys. Lett. B",
    volume = "729",
    pages = "108--111",
    year = "2014"
}

@article{Aarts:2014bwa,
    author = "Aarts, Gert and Seiler, Erhard and Sexty, D{\'e}nes and Stamatescu, Ion-Olimpiu",
    title = "{Simulating QCD at nonzero baryon density to all orders in the hopping parameter expansion}",
    eprint = "1408.3770",
    archivePrefix = "arXiv",
    primaryClass = "hep-lat",
    doi = "10.1103/PhysRevD.90.114505",
    journal = "Phys. Rev. D",
    volume = "90",
    number = "11",
    pages = "114505",
    year = "2014"
}

@article{Fodor:2015doa,
    author = {Fodor, Z. and Katz, S. D. and Sexty, D. and T{\"o}r{\"o}k, C.},
    title = "{Complex Langevin dynamics for dynamical QCD at nonzero chemical potential: A comparison with multiparameter reweighting}",
    eprint = "1508.05260",
    archivePrefix = "arXiv",
    primaryClass = "hep-lat",
    doi = "10.1103/PhysRevD.92.094516",
    journal = "Phys. Rev. D",
    volume = "92",
    number = "9",
    pages = "094516",
    year = "2015"
}

@article{Nagata:2018mkb,
    author = "Nagata, Keitaro and Nishimura, Jun and Shimasaki, Shinji",
    title = "{Complex Langevin calculations in finite density QCD at large {\ensuremath{\mu}}/T with the deformation technique}",
    eprint = "1805.03964",
    archivePrefix = "arXiv",
    primaryClass = "hep-lat",
    reportNumber = "KEK-TH-2048",
    doi = "10.1103/PhysRevD.98.114513",
    journal = "Phys. Rev. D",
    volume = "98",
    number = "11",
    pages = "114513",
    year = "2018"
}

@article{Kogut:2019qmi,
    author = "Kogut, J. B. and Sinclair, D. K.",
    title = "{Applying Complex Langevin Simulations to Lattice QCD at Finite Density}",
    eprint = "1903.02622",
    archivePrefix = "arXiv",
    primaryClass = "hep-lat",
    doi = "10.1103/PhysRevD.100.054512",
    journal = "Phys. Rev. D",
    volume = "100",
    number = "5",
    pages = "054512",
    year = "2019"
}

@article{Sexty:2019vqx,
    author = "Sexty, D{\'e}nes",
    title = "{Calculating the equation of state of dense quark-gluon plasma using the complex Langevin equation}",
    eprint = "1907.08712",
    archivePrefix = "arXiv",
    primaryClass = "hep-lat",
    doi = "10.1103/PhysRevD.100.074503",
    journal = "Phys. Rev. D",
    volume = "100",
    number = "7",
    pages = "074503",
    year = "2019"
}

@inproceedings{Longia:2026doi,
    author = "Longia, Vamika and Dhindsa, Navdeep Singh and Joseph, Anosh",
    title = "{Configurational Thermometer for Lattice Gauge Theories}",
    booktitle = "{42th International Symposium on Lattice Field Theory}",
    eprint = "2601.17436",
    archivePrefix = "arXiv",
    primaryClass = "hep-lat",
    reportNumber = "TIFR/TH/26-6",
    month = "1",
    year = "2026"
}

@article{Dhindsa:2025xfv,
    author = "Dhindsa, Navdeep Singh and Joseph, Anosh and Longia, Vamika",
    title = "{Gradient and Hessian-Based temperature estimator in lattice gauge theories: a diagnostic tool for stability and consistency in numerical simulations}",
    eprint = "2508.05595",
    archivePrefix = "arXiv",
    primaryClass = "hep-lat",
    doi = "10.1007/JHEP10(2025)015",
    journal = "JHEP",
    volume = "10",
    pages = "015",
    year = "2025"
}

@article{Joseph:2025xbn,
    author = "Joseph, Anosh and Kumar, Arpith",
    title = "{Configurational Temperature as a Diagnostic for Complex Langevin Dynamics in the 3D XY Model}",
    eprint = "2509.13314",
    archivePrefix = "arXiv",
    primaryClass = "hep-lat",
    month = "9",
    year = "2025"
}

@article{Mandl:2025mav,
    author = "Mandl, Michael and Seiler, Erhard and Sexty, D{\'e}nes",
    title = "{Necessary and sufficient conditions for correctness of complex Langevin}",
    eprint = "2508.14512",
    archivePrefix = "arXiv",
    primaryClass = "hep-lat",
    month = "8",
    year = "2025"
}

@article{Scherzer:2019lrh,
    author = "Scherzer, M. and Seiler, E. and Sexty, D. and Stamatescu, I. -O.",
    title = "{Controlling Complex Langevin simulations of lattice models by boundary term analysis}",
    eprint = "1910.09427",
    archivePrefix = "arXiv",
    primaryClass = "hep-lat",
    doi = "10.1103/PhysRevD.101.014501",
    journal = "Phys. Rev. D",
    volume = "101",
    number = "1",
    pages = "014501",
    year = "2020"
}

@article{Seiler:2023kes,
    author = "Seiler, Erhard and Sexty, D{\'e}nes and Stamatescu, Ion-Olimpiu",
    title = "{Complex Langevin: Correctness criteria, boundary terms, and spectrum}",
    eprint = "2304.00563",
    archivePrefix = "arXiv",
    primaryClass = "hep-lat",
    doi = "10.1103/PhysRevD.109.014509",
    journal = "Phys. Rev. D",
    volume = "109",
    number = "1",
    pages = "014509",
    year = "2024"
}

@article{Scherzer:2018hid,
    author = "Scherzer, Manuel and Seiler, Erhard and Sexty, D{\'e}nes and Stamatescu, Ion-Olimpiu",
    title = "{Complex Langevin and boundary terms}",
    eprint = "1808.05187",
    archivePrefix = "arXiv",
    primaryClass = "hep-lat",
    doi = "10.1103/PhysRevD.99.014512",
    journal = "Phys. Rev. D",
    volume = "99",
    number = "1",
    pages = "014512",
    year = "2019"
}

@article{Uhlenbeck:1930zz,
    author = "Uhlenbeck, G. E. and Ornstein, L. S.",
    title = "{On the Theory of the Brownian Motion}",
    doi = "10.1103/PhysRev.36.823",
    journal = "Phys. Rev.",
    volume = "36",
    pages = "823--841",
    year = "1930"
}

@article{Kubo:1966fyg,
    author = "Kubo, R.",
    title = "{The fluctuation-dissipation theorem}",
    doi = "10.1088/0034-4885/29/1/306",
    journal = "Rept. Prog. Phys.",
    volume = "29",
    number = "1",
    pages = "255",
    year = "1966"
}

@article{Bender:1998ke,
    author = "Bender, Carl M. and Boettcher, Stefan",
    title = "{Real spectra in nonHermitian Hamiltonians having PT symmetry}",
    eprint = "physics/9712001",
    archivePrefix = "arXiv",
    doi = "10.1103/PhysRevLett.80.5243",
    journal = "Phys. Rev. Lett.",
    volume = "80",
    pages = "5243--5246",
    year = "1998"
}

@article{Bender:1998gh,
    author = "Bender, Carl M. and Boettcher, Stefan and Meisinger, Peter",
    title = "{PT symmetric quantum mechanics}",
    eprint = "quant-ph/9809072",
    archivePrefix = "arXiv",
    doi = "10.1063/1.532860",
    journal = "J. Math. Phys.",
    volume = "40",
    pages = "2201--2229",
    year = "1999"
}

@article{Seiler:2017wvd,
    author = "Seiler, Erhard",
    editor = "Della Morte, M. and Fritzsch, P. and G{\'a}miz S{\'a}nchez, E. and Pena Ruano, C.",
    title = "{Status of Complex Langevin}",
    eprint = "1708.08254",
    archivePrefix = "arXiv",
    primaryClass = "hep-lat",
    reportNumber = "MPP-2017-182",
    doi = "10.1051/epjconf/201817501019",
    journal = "EPJ Web Conf.",
    volume = "175",
    pages = "01019",
    year = "2018"
}

@article{Aarts:2017vrv,
    author = "Aarts, Gert and Seiler, Erhard and Sexty, Denes and Stamatescu, Ion-Olimpiu",
    title = "{Complex Langevin dynamics and zeroes of the fermion determinant}",
    eprint = "1701.02322",
    archivePrefix = "arXiv",
    primaryClass = "hep-lat",
    doi = "10.1007/JHEP05(2017)044",
    journal = "JHEP",
    volume = "05",
    pages = "044",
    year = "2017",
    note = "[Erratum: JHEP 01, 128 (2018)]"
}

@article{Nishimura:2015pba,
    author = "Nishimura, Jun and Shimasaki, Shinji",
    title = "{New Insights into the Problem with a Singular Drift Term in the Complex Langevin Method}",
    eprint = "1504.08359",
    archivePrefix = "arXiv",
    primaryClass = "hep-lat",
    reportNumber = "KEK-TH-1816",
    doi = "10.1103/PhysRevD.92.011501",
    journal = "Phys. Rev. D",
    volume = "92",
    number = "1",
    pages = "011501",
    year = "2015"
}

@article{Joseph:2019sof,
    author = "Joseph, Anosh and Kumar, Arpith",
    title = "{Complex Langevin Simulations of Zero-dimensional Supersymmetric Quantum Field Theories}",
    eprint = "1908.04153",
    archivePrefix = "arXiv",
    primaryClass = "hep-th",
    doi = "10.1103/PhysRevD.100.074507",
    journal = "Phys. Rev. D",
    volume = "100",
    pages = "074507",
    year = "2019"
}

@article{Joseph:2020gdh,
    author = "Joseph, Anosh and Kumar, Arpith",
    title = "{Complex Langevin dynamics and supersymmetric quantum mechanics}",
    eprint = "2011.08107",
    archivePrefix = "arXiv",
    primaryClass = "hep-lat",
    doi = "10.1007/JHEP10(2021)186",
    journal = "JHEP",
    volume = "10",
    pages = "186",
    year = "2021"
}

@article{Kumar:2022fas,
    author = "Kumar, Arpith and Joseph, Anosh",
    title = "{Complex Langevin simulations for PT-symmetric models}",
    eprint = "2201.12001",
    archivePrefix = "arXiv",
    primaryClass = "hep-lat",
    doi = "10.22323/1.396.0124",
    journal = "PoS",
    volume = "LATTICE2021",
    pages = "124",
    year = "2022"
}

@article{Kumar:2022giw,
    author = "Kumar, Arpith and Joseph, Anosh and Kumar, Piyush",
    title = "{Complex Langevin Study of Spontaneous Symmetry Breaking in IKKT Matrix Model}",
    eprint = "2209.10494",
    archivePrefix = "arXiv",
    primaryClass = "hep-lat",
    doi = "10.22323/1.430.0213",
    journal = "PoS",
    volume = "LATTICE2022",
    pages = "213",
    year = "2023"
}

@article{Kumar:2023nya,
    author = "Kumar, Arpith and Joseph, Anosh and Kumar, Piyush",
    title = "{Investigating Spontaneous SO(10) Symmetry Breaking in~Type IIB Matrix Model}",
    eprint = "2308.03607",
    archivePrefix = "arXiv",
    primaryClass = "hep-lat",
    doi = "10.1007/978-981-97-0289-3_337",
    journal = "Springer Proc. Phys.",
    volume = "304",
    pages = "1201--1203",
    year = "2024"
}

@article{Joseph:2025tfw,
    author = "Joseph, Anosh and Kumar, Arpith",
    title = "{Complex Langevin simulations of supersymmetric theories}",
    eprint = "2504.02660",
    archivePrefix = "arXiv",
    primaryClass = "hep-lat",
    doi = "10.1142/S0217751X25300066",
    journal = "Int. J. Mod. Phys. A",
    volume = "40",
    number = "20",
    pages = "2530006",
    year = "2025"
}

@article{Berger:2019odf,
    author = {Berger, Casey E. and Rammelm{\"u}ller, Lukas and Loheac, Andrew C. and Ehmann, Florian and Braun, Jens and Drut, Joaqu{\'\i}n E.},
    title = "{Complex Langevin and other approaches to the sign problem in quantum many-body physics}",
    eprint = "1907.10183",
    archivePrefix = "arXiv",
    primaryClass = "cond-mat.quant-gas",
    doi = "10.1016/j.physrep.2020.09.002",
    journal = "Phys. Rept.",
    volume = "892",
    pages = "1--54",
    year = "2021"
}

@article{Klauder:1983nn,
author         = "Klauder, J. R.",
title          = "{Stochastic Quantization}",
booktitle      = "{Recent developments in high-energy physics. Proceedings,
22. Internationale Universitatswochen fur Kernphysik:
Schladming, Austria, February 23 - March 5, 1983}",
journal        = "Acta Phys. Austriaca Suppl.",
volume         = "25",
year           = "1983",
pages          = "251-281",
doi            = "10.1007/978-3-7091-7651-1_8",
reportNumber   = "PRINT-83-0321 (BTL)",
SLACcitation   = "%%CITATION = APAUA,25,251;%%"
}

@article{Klauder:1983zm,
author         = "Klauder, John R.",
title          = "{A Langevin Approach to Fermion and Quantum Spin
Correlation Functions}",
journal        = "J. Phys.",
volume         = "A16",
year           = "1983",
pages          = "L317",
doi            = "10.1088/0305-4470/16/10/001",
reportNumber   = "Print-83-0008 (BTL)",
SLACcitation   = "%%CITATION = JPAGA,A16,L317;%%"
}

@article{Klauder:1983sp,
author         = "Klauder, John R.",
title          = "{Coherent State Langevin Equations for Canonical Quantum
Systems With Applications to the Quantized Hall Effect}",
journal        = "Phys. Rev.",
volume         = "A29",
year           = "1984",
pages          = "2036-2047",
doi            = "10.1103/PhysRevA.29.2036",
reportNumber   = "Print-83-0902 (BTL)",
SLACcitation   = "%%CITATION = PHRVA,A29,2036;%%"
}

@article{Parisi:1984cs,
author         = "Parisi, G.",
title          = "{On Complex Probabilities}",
journal        = "Phys. Lett.",
volume         = "131B",
year           = "1983",
pages          = "393-395",
doi            = "10.1016/0370-2693(83)90525-7",
SLACcitation   = "%%CITATION = PHLTA,131B,393;%%"
}

@article{Damgaard:1987rr,
author         = "Damgaard, Poul H. and Huffel, Helmuth",
title          = "{Stochastic Quantization}",
journal        = "Phys. Rept.",
volume         = "152",
year           = "1987",
pages          = "227",
doi            = "10.1016/0370-1573(87)90144-X",
reportNumber   = "PRINT-87-0157 (CERN)",
SLACcitation   = "%%CITATION = PRPLC,152,227;%%"
}

@article{Basu:2018dtm,
author         = "Basu, Pallab and Jaswin, Kasi and Joseph, Anosh",
title          = "{Complex Langevin Dynamics in Large $N$ Unitary Matrix
Models}",
journal        = "Phys. Rev.",
volume         = "D98",
year           = "2018",
number         = "3",
pages          = "034501",
doi            = "10.1103/PhysRevD.98.034501",
eprint         = "1802.10381",
archivePrefix  = "arXiv",
primaryClass   = "hep-th",
SLACcitation   = "%%CITATION = ARXIV:1802.10381;%%"
}

@article{Gross:1980he,
author         = "Gross, D. J. and Witten, Edward",
title          = "{Possible Third Order Phase Transition in the Large N
Lattice Gauge Theory}",
journal        = "Phys. Rev.",
volume         = "D21",
year           = "1980",
pages          = "446-453",
doi            = "10.1103/PhysRevD.21.446",
SLACcitation   = "%%CITATION = PHRVA,D21,446;%%"
}

@article{Wadia:1980cp,
author         = "Wadia, Spenta R.",
title          = "{$N$ = Infinity Phase Transition in a Class of Exactly
Soluble Model Lattice Gauge Theories}",
journal        = "Phys. Lett.",
volume         = "93B",
year           = "1980",
pages          = "403-410",
doi            = "10.1016/0370-2693(80)90353-6",
reportNumber   = "EFI-80/15-CHICAGO",
SLACcitation   = "%%CITATION = PHLTA,93B,403;%%"
}

@article{Wadia:2012fr,
author         = "Wadia, Spenta R.",
title          = "{A Study of U(N) Lattice Gauge Theory in 2-dimensions}",
year           = "2012",
eprint         = "1212.2906",
archivePrefix  = "arXiv",
primaryClass   = "hep-th",
reportNumber   = "ICTS-2012-13, TIFR-TH-2012-47",
SLACcitation   = "%%CITATION = ARXIV:1212.2906;%%"
}

@article{Berges:2005yt,
author         = "Berges, J. and Stamatescu, I. -O.",
title          = "{Simulating nonequilibrium quantum fields with stochastic
quantization techniques}",
journal        = "Phys. Rev. Lett.",
volume         = "95",
year           = "2005",
pages          = "202003",
doi            = "10.1103/PhysRevLett.95.202003",
eprint         = "hep-lat/0508030",
archivePrefix  = "arXiv",
primaryClass   = "hep-lat",
SLACcitation   = "%%CITATION = HEP-LAT/0508030;%%"
}

@article{Berges:2006xc,
author         = "Berges, J. and Borsanyi, Sz. and Sexty, D. and
Stamatescu, I. -O.",
title          = "{Lattice simulations of real-time quantum fields}",
journal        = "Phys. Rev.",
volume         = "D75",
year           = "2007",
pages          = "045007",
doi            = "10.1103/PhysRevD.75.045007",
eprint         = "hep-lat/0609058",
archivePrefix  = "arXiv",
primaryClass   = "hep-lat",
SLACcitation   = "%%CITATION = HEP-LAT/0609058;%%"
}

@article{Berges:2007nr,
author         = "Berges, Juergen and Sexty, Denes",
title          = "{Real-time gauge theory simulations from stochastic
quantization with optimized updating}",
journal        = "Nucl. Phys.",
volume         = "B799",
year           = "2008",
pages          = "306-329",
doi            = "10.1016/j.nuclphysb.2008.01.018",
eprint         = "0708.0779",
archivePrefix  = "arXiv",
primaryClass   = "hep-lat",
SLACcitation   = "%%CITATION = ARXIV:0708.0779;%%"
}

@article{Bloch:2017sex,
author         = "Bloch, J. and Glesaaen, J. and Verbaarschot, J. J. M. and
Zafeiropoulos, S.",
title          = "{Complex Langevin Simulation of a Random Matrix Model at
Nonzero Chemical Potential}",
journal        = "JHEP",
volume         = "03",
year           = "2018",
pages          = "015",
doi            = "10.1007/JHEP03(2018)015",
eprint         = "1712.07514",
archivePrefix  = "arXiv",
primaryClass   = "hep-lat",
SLACcitation   = "%%CITATION = ARXIV:1712.07514;%%"
}

@article{Aarts:2008rr,
author         = "Aarts, Gert and Stamatescu, Ion-Olimpiu",
title          = "{Stochastic quantization at finite chemical potential}",
journal        = "JHEP",
volume         = "09",
year           = "2008",
pages          = "018",
doi            = "10.1088/1126-6708/2008/09/018",
eprint         = "0807.1597",
archivePrefix  = "arXiv",
primaryClass   = "hep-lat",
SLACcitation   = "%%CITATION = ARXIV:0807.1597;%%"
}

@article{Pehlevan:2007eq,
author         = "Pehlevan, Cengiz and Guralnik, Gerald",
title          = "{Complex Langevin Equations and Schwinger-Dyson
Equations}",
journal        = "Nucl. Phys.",
volume         = "B811",
year           = "2009",
pages          = "519-536",
doi            = "10.1016/j.nuclphysb.2008.11.034",
eprint         = "0710.3756",
archivePrefix  = "arXiv",
primaryClass   = "hep-th",
reportNumber   = "BROWN-HET-1488",
SLACcitation   = "%%CITATION = ARXIV:0710.3756;%%"
}

@article{Aarts:2008wh,
author         = "Aarts, Gert",
title          = "{Can stochastic quantization evade the sign problem? The
relativistic Bose gas at finite chemical potential}",
journal        = "Phys. Rev. Lett.",
volume         = "102",
year           = "2009",
pages          = "131601",
doi            = "10.1103/PhysRevLett.102.131601",
eprint         = "0810.2089",
archivePrefix  = "arXiv",
primaryClass   = "hep-lat",
SLACcitation   = "%%CITATION = ARXIV:0810.2089;%%"
}

@article{Aarts:2009hn,
author         = "Aarts, Gert",
title          = "{Complex Langevin dynamics at finite chemical potential:
Mean field analysis in the relativistic Bose gas}",
journal        = "JHEP",
volume         = "05",
year           = "2009",
pages          = "052",
doi            = "10.1088/1126-6708/2009/05/052",
eprint         = "0902.4686",
archivePrefix  = "arXiv",
primaryClass   = "hep-lat",
SLACcitation   = "%%CITATION = ARXIV:0902.4686;%%"
}

@article{Aarts:2010gr,
author         = "Aarts, Gert and Splittorff, K.",
title          = "{Degenerate distributions in complex Langevin dynamics:
one-dimensional QCD at finite chemical potential}",
journal        = "JHEP",
volume         = "08",
year           = "2010",
pages          = "017",
doi            = "10.1007/JHEP08(2010)017",
eprint         = "1006.0332",
archivePrefix  = "arXiv",
primaryClass   = "hep-lat",
SLACcitation   = "%%CITATION = ARXIV:1006.0332;%%"
}

@article{Aarts:2011zn,
author         = "Aarts, Gert and James, Frank A.",
title          = "{Complex Langevin dynamics in the SU(3) spin model at
nonzero chemical potential revisited}",
journal        = "JHEP",
volume         = "01",
year           = "2012",
pages          = "118",
doi            = "10.1007/JHEP01(2012)118",
eprint         = "1112.4655",
archivePrefix  = "arXiv",
primaryClass   = "hep-lat",
SLACcitation   = "%%CITATION = ARXIV:1112.4655;%%"
}

@article{Ito:2016efb,
author         = "Ito, Yuta and Nishimura, Jun",
title          = "{The complex Langevin analysis of spontaneous symmetry
breaking induced by complex fermion determinant}",
journal        = "JHEP",
volume         = "12",
year           = "2016",
pages          = "009",
doi            = "10.1007/JHEP12(2016)009",
eprint         = "1609.04501",
archivePrefix  = "arXiv",
primaryClass   = "hep-lat",
reportNumber   = "KEK-TH-1934",
SLACcitation   = "%%CITATION = ARXIV:1609.04501;%%"
}

@article{Ito:2016hlj,
author         = "Ito, Yuta and Nishimura, Jun",
title          = "{Spontaneous symmetry breaking induced by complex fermion
determinant --- yet another success of the complex
Langevin method}",
booktitle      = "{Proceedings, 34th International Symposium on Lattice
Field Theory (Lattice 2016): Southampton, UK, July 24-30,
2016}",
journal        = "PoS",
volume         = "LATTICE2016",
year           = "2016",
pages          = "065",
doi            = "10.22323/1.256.0065",
eprint         = "1612.00598",
archivePrefix  = "arXiv",
primaryClass   = "hep-lat",
reportNumber   = "KEK-TH-1948",
SLACcitation   = "%%CITATION = ARXIV:1612.00598;%%"
}

@article{Anagnostopoulos:2017gos,
author         = "Anagnostopoulos, Konstantinos N. and Azuma, Takehiro and
Ito, Yuta and Nishimura, Jun and Papadoudis, Stratos
Kovalkov",
title          = "{Complex Langevin analysis of the spontaneous symmetry
breaking in dimensionally reduced super Yang-Mills
models}",
journal        = "JHEP",
volume         = "02",
year           = "2018",
pages          = "151",
doi            = "10.1007/JHEP02(2018)151",
eprint         = "1712.07562",
archivePrefix  = "arXiv",
primaryClass   = "hep-lat",
reportNumber   = "KEK-TH-2023",
SLACcitation   = "%%CITATION = ARXIV:1712.07562;%%"
}

@article{Aarts:2013uza,
author         = "Aarts, Gert and Giudice, Pietro and Seiler, Erhard",
title          = "{Localised distributions and criteria for correctness in
complex Langevin dynamics}",
journal        = "Annals Phys.",
volume         = "337",
year           = "2013",
pages          = "238-260",
doi            = "10.1016/j.aop.2013.06.019",
eprint         = "1306.3075",
archivePrefix  = "arXiv",
primaryClass   = "hep-lat",
SLACcitation   = "%%CITATION = ARXIV:1306.3075;%%"
}

@article{Nagata:2015uga,
author         = "Nagata, Keitaro and Nishimura, Jun and Shimasaki, Shinji",
title          = "{Justification of the complex Langevin method with the
gauge cooling procedure}",
journal        = "PTEP",
volume         = "2016",
year           = "2016",
number         = "1",
pages          = "013B01",
doi            = "10.1093/ptep/ptv173",
eprint         = "1508.02377",
archivePrefix  = "arXiv",
primaryClass   = "hep-lat",
reportNumber   = "KEK-TH-1855",
SLACcitation   = "%%CITATION = ARXIV:1508.02377;%%"
}

@article{Aarts:2009uq,
author         = "Aarts, Gert and Seiler, Erhard and Stamatescu,
Ion-Olimpiu",
title          = "{The Complex Langevin method: When can it be trusted?}",
journal        = "Phys. Rev.",
volume         = "D81",
year           = "2010",
pages          = "054508",
doi            = "10.1103/PhysRevD.81.054508",
eprint         = "0912.3360",
archivePrefix  = "arXiv",
primaryClass   = "hep-lat",
SLACcitation   = "%%CITATION = ARXIV:0912.3360;%%"
}

@article{Aarts:2011ax,
author         = "Aarts, Gert and James, Frank A. and Seiler, Erhard and
Stamatescu, Ion-Olimpiu",
title          = "{Complex Langevin: Etiology and Diagnostics of its Main
Problem}",
journal        = "Eur. Phys. J.",
volume         = "C71",
year           = "2011",
pages          = "1756",
doi            = "10.1140/epjc/s10052-011-1756-5",
eprint         = "1101.3270",
archivePrefix  = "arXiv",
primaryClass   = "hep-lat",
reportNumber   = "MPP-2011-3",
SLACcitation   = "%%CITATION = ARXIV:1101.3270;%%"
}

@article{Bender:1999ek,
author         = "Bender, Carl M. and Milton, Kimball A. and Savage, Van
M.",
title          = "{Solution of Schwinger-Dyson equations for PT symmetric
quantum field theory}",
journal        = "Phys. Rev.",
volume         = "D62",
year           = "2000",
pages          = "085001",
doi            = "10.1103/PhysRevD.62.085001",
eprint         = "hep-th/9907045",
archivePrefix  = "arXiv",
primaryClass   = "hep-th",
reportNumber   = "OKHEP-99-05",
SLACcitation   = "%%CITATION = HEP-TH/9907045;%%"
}

@article{Nagata:2016vkn,
author         = "Nagata, Keitaro and Nishimura, Jun and Shimasaki, Shinji",
title          = "{Argument for justification of the complex Langevin
method and the condition for correct convergence}",
journal        = "Phys. Rev.",
volume         = "D94",
year           = "2016",
number         = "11",
pages          = "114515",
doi            = "10.1103/PhysRevD.94.114515",
eprint         = "1606.07627",
archivePrefix  = "arXiv",
primaryClass   = "hep-lat",
reportNumber   = "KEK-TH-1911",
SLACcitation   = "%%CITATION = ARXIV:1606.07627;%%"
}

@article{Nagata:2018net,
author         = "Nagata, Keitaro and Nishimura, Jun and Shimasaki, Shinji",
title          = "{Testing the criterion for correct convergence in the
complex Langevin method}",
journal        = "JHEP",
volume         = "05",
year           = "2018",
pages          = "004",
doi            = "10.1007/JHEP05(2018)004",
eprint         = "1802.01876",
archivePrefix  = "arXiv",
primaryClass   = "hep-lat",
reportNumber   = "KEK-TH-2032",
SLACcitation   = "%%CITATION = ARXIV:1802.01876;%%"
}

@article{Klauder:1985kq,
author         = "Klauder, John R. and Petersen, Wesley P.",
title          = "{Numerical Integration of Multiplicative Noise Stochastic Differential Equations}",
journal        = "SIAM J. Num. Anal.",
volume         = "22",
year           = "1985",
pages          = "1153-1166",
reportNumber   = "PRINT-85-0297 (BTL)",
SLACcitation   = "%%CITATION = PRINT-85-0297 (BTL);%%"
}

@article{Klauder:1985ks,
author         = "Klauder, John R. and Petersen, Wesley P.",
title          = "{Spectrum of Certain Non-self-adjoint Operators and Solutions of Langevin Equations with Complex Drift}",
journal        = "J. Stat. Phys.",
volume         = "39",
year           = "1985",
pages          = "53-72",
reportNumber   = "Print-85-0295 (BTL)",
SLACcitation   = "%%CITATION = PRINT-85-0295 (BTL);%%"
}

@article{Gausterer:1986gk,
author         = "Gausterer, H. and Klauder, J. R.",
title          = "{Complex Langevin Equations and Their Applications to
Quantum Statistical and Lattice Field Models}",
journal        = "Phys. Rev.",
volume         = "D33",
year           = "1986",
pages          = "3678",
doi            = "10.1103/PhysRevD.33.3678",
reportNumber   = "UNIGRAZ-UTP-1-86",
SLACcitation   = "%%CITATION = PHRVA,D33,3678;%%"
}

@article{10.1063/1.480995,
    author = {Baranyai, Andras},
    title = {On the configurational temperature of simple fluids},
    journal = {The Journal of Chemical Physics},
    volume = {112},
    number = {9},
    pages = {3964-3966},
    year = {2000},
    month = {03},
    issn = {0021-9606},
    doi = {10.1063/1.480995},
    url = {https://doi.org/10.1063/1.480995},
    eprint = {https://pubs.aip.org/aip/jcp/article-pdf/112/9/3964/19250367/3964\_1\_online.pdf},
}

@article{10.1063/1.1348024,
    author = {Rickayzen, Gerald and Powles, Jack G.},
    title = {Temperature in the classical microcanonical ensemble},
    journal = {The Journal of Chemical Physics},
    volume = {114},
    number = {9},
    pages = {4333-4334},
    year = {2001},
    month = {03},
    issn = {0021-9606},
    doi = {10.1063/1.1348024},
    url = {https://doi.org/10.1063/1.1348024},
    eprint = {https://pubs.aip.org/aip/jcp/article-pdf/114/9/4333/19036407/4333\_1\_online.pdf},
}

@ARTICLE{1998JPhA...31.7761R,
       author = {{Rugh}, Hans Henrik},
        title = "{A geometric, dynamical approach to thermodynamics}",
      journal = {Journal of Physics A Mathematical General},
         year = 1998,
        month = sep,
       volume = {31},
       number = {38},
        pages = {7761-7770},
          doi = {10.1088/0305-4470/31/38/011},
archivePrefix = {arXiv},
       eprint = {chao-dyn/9703013},
 primaryClass = {nlin.CD},
       adsurl = {https://ui.adsabs.harvard.edu/abs/1998JPhA...31.7761R}
}

@article{PhysRevE.94.062113,
  title = {Ensemble-free configurational temperature for spin systems},
  author = {Palma, G. and Guti\'errez, G. and Davis, S.},
  journal = {Phys. Rev. E},
  volume = {94},
  issue = {6},
  pages = {062113},
  numpages = {8},
  year = {2016},
  month = {Dec},
  publisher = {American Physical Society},
  doi = {10.1103/PhysRevE.94.062113},
  url = {https://link.aps.org/doi/10.1103/PhysRevE.94.062113}
}

@article{PhysRevE.62.4757,
  title = {Microscopic expressions for the thermodynamic temperature},
  author = {Jepps, Owen G. and Ayton, Gary and Evans, Denis J.},
  journal = {Phys. Rev. E},
  volume = {62},
  issue = {4},
  pages = {4757--4763},
  numpages = {0},
  year = {2000},
  month = {Oct},
  publisher = {American Physical Society},
  doi = {10.1103/PhysRevE.62.4757},
  url = {https://link.aps.org/doi/10.1103/PhysRevE.62.4757}
}

@article{PhysRevLett.78.772,
  title = {Dynamical Approach to Temperature},
  author = {Rugh, Hans Henrik},
  journal = {Phys. Rev. Lett.},
  volume = {78},
  issue = {5},
  pages = {772--774},
  numpages = {0},
  year = {1997},
  month = {Feb},
  eprint = "chao-dyn/9701026",
  archivePrefix = "arXiv",
  publisher = {American Physical Society},
  doi = {10.1103/PhysRevLett.78.772},
  url = {https://link.aps.org/doi/10.1103/PhysRevLett.78.772}
}

@article{10.1063/1.477301,
    author = {Butler, B. D. and Ayton, Gary and Jepps, Owen G. and Evans, Denis J.},
    title = {Configurational temperature: Verification of Monte Carlo simulations},
    journal = {The Journal of Chemical Physics},
    volume = {109},
    number = {16},
    pages = {6519-6522},
    year = {1998},
    month = {10},
    issn = {0021-9606},
    doi = {10.1063/1.477301},
    url = {https://doi.org/10.1063/1.477301}
}

@book{Landau1952,
  author    = {L. D. Landau and E. M. Lifshitz},
  title     = {Statisticheskaya Fizika (Statistical Physics)},
  year      = {1952},
  publisher = {Nauka},
  address   = {Moscow},
  note      = {See Eq. 33.14}
}

@book{Tolman1979,
  author    = {R. C. Tolman},
  title     = {The Principles of Statistical Mechanics},
  year      = {1979},
  publisher = {Dover},
  address   = {New York}
}

@article{Hoover2007,
  author  = {W. G. Hoover and C. G. Hoover},
  title   = {Hamiltonian dynamics of thermostated systems: Two-temperature heat-conducting $\phi^4$ chains},
  journal = {Journal of Chemical Physics},
  volume  = {126},
  pages   = {164113},
  year    = {2007},
  doi     = {10.1063/1.2720839}
}

@article{Bernard:2001wh,
    author = "Bernard, Claude W. and Savage, Van M.",
    title = "{Numerical simulations of PT symmetric quantum field theories}",
    eprint = "hep-lat/0106009",
    archivePrefix = "arXiv",
    doi = "10.1103/PhysRevD.64.085010",
    journal = "Phys. Rev. D",
    volume = "64",
    pages = "085010",
    year = "2001"
}
\end{document}